\documentclass[final,3p,times,numbers,sort&compress]{elsarticle}

\usepackage{amsmath}
\usepackage[utf8]{inputenc}
\usepackage{bm}
\usepackage{amsfonts}
\usepackage{amsthm}
\usepackage{amssymb}
\usepackage{mathrsfs}
\usepackage{graphicx}

\journal{Physica A}

\usepackage[breaklinks=true,colorlinks=true,linkcolor=blue,urlcolor=blue,citecolor=blue]{hyperref}

\renewcommand{\eqref}[1]{(\ref{#1})}
\newcommand{\defeq}{\mathrel{\mathop:}=}
\newcommand{\U}{\mathcal{U}}
\newcommand{\D}{\mathcal{D}}
\renewcommand{\H}{\mathcal{H}}

\newcommand{\vth}{v_{\text{th}}}
\newcommand{\params}{\boldsymbol{\lambda}}

\begin{document}

\begin{frontmatter}

\title{Kappa distributions in the framework of superstatistics}

\author[cchen,unab]{Sergio Davis\corref{cor1}}
\ead{sergio.davis@cchen.cl}

\author[cchen,unab]{Biswajit Bora}
\author[cchen,unab]{Cristian Pavez}
\author[cchen,unab]{Leopoldo Soto}

\address[cchen]{Research Center on the Intersection of Plasma Physics, Matter and Complexity (P$^2$mc),\\ Comisión Chilena de Energía Nuclear, Casilla 188-D, Santiago, Chile}
\address[unab]{Departamento de Física y Astronomía, Facultad de Ciencias Exactas, Universidad Andres Bello,\\ Sazié 2212, piso 7, 8370136, Santiago, Chile.}

\cortext[cor1]{Corresponding author}

\begin{abstract}
The kappa distribution of velocities is frequently found instead of the Maxwellian distribution in collisionless plasmas present in Earth's magnetosphere, the solar wind among other contexts 
where particles do not reach thermal equilibrium. Although the origin of these distributions is sometimes explained by means of non-extensive statistics, they can also be recovered using alternative 
frameworks such as superstatistics, providing a closer connection with probability theory. In this work we take this approach and derive the multi-particle and single-particle kappa distributions 
from superstatistics while taking into account the scale invariance property of the superstatistical temperature distribution. The formalism presented here emphasizes the usefulness of superstatistics 
in the computation of expectation values under kappa distributions. Some consequences of a superstatistical interpretation of kappa distributions are also discussed, such as the connection between 
correlations and temperature uncertainty, the meaning of the superstatistical temperature and the entropy of kappa-distributed plasmas.
\end{abstract}

\end{frontmatter}

\section{Introduction}

As is well known, systems of particles in thermal equilibrium have velocities described by the Maxwell-Boltzmann distribution. For a particle with mass $m$ at temperature 
$T$, the Maxwell-Boltzmann distribution is the three-dimensional Gaussian
\begin{equation}
\label{eq:maxwell}
P(\bm v|\beta, m) = \left(\frac{m\beta}{2\pi}\right)^{\frac{3}{2}} \exp\left(-\frac{\beta m\bm{v}^2}{2}\right),
\end{equation}
where
\begin{equation}
\beta \defeq \frac{1}{k_B T}
\end{equation}
is the inverse temperature. On the other hand, there are systems such as collisionless plasmas present in the Earth's magnetosphere~\cite{Antonova2008, Espinoza2018, Kirpichev2020, Eyelade2021}, the solar wind~\cite{Maksimovic1997b, Nicolaou2019, ZentenoQuinteros2021} and the interstellar medium~\cite{Raymond2010, Nicholls2017} which are unable to reach thermal equilibrium, and in that case what is commonly 
observed is that particle velocities are actually described by non-Maxwellian distributions, generalizations of \eqref{eq:maxwell}. Among them the kappa distribution~\cite{Pierrard2010, Livadiotis2017, Lazar2021} is 
probably the most common, although other models such as the Cairns distribution~\cite{Cairns1995, Naim2019, Khalid2025}, the Kaniadakis distribution~\cite{Kaniadakis2001, Lourek2016, Gougam2016, Lopez2017b} and the 
super-Gaussian~\cite{Dum1974, Bissell2013} distribution are also used.

Outside space plasmas, kappa distributions are also plausible models for the velocity distributions in laboratory plasmas such as Z-pinch discharges, but have only recently been proposed~\cite{Knapp2013, Klir2015}. 
As these kind of plasma discharges have already been characterized as presenting power laws~\cite{Potter1971,Bernstein1972,Stygar1982,Vikhrev2007,Vikhrev2012} and phenomena such as jets and filaments similar to the 
ones in astrophysical environments have been observed~\cite{Soto2014, Pavez2015, Pavez2022}, an interesting possibility is the existence of kappa-distributed velocities.  Under one common 
parameterization~\cite{Livadiotis2017}, the kappa distribution is written as
\begin{equation}
\label{eq:kappa}
P(\bm v|\kappa, \vth) = \left[\pi\left(\kappa-\frac{3}{2}\right) \vth^2\right]^{-\frac{3}{2}} \frac{\Gamma(\kappa+1)}{\Gamma(\kappa-\frac{1}{2})}
\left[1 + \frac{1}{\kappa-\frac{3}{2}}\frac{\bm{v}^2}{\vth^2}\right]^{-(\kappa+1)},
\end{equation}
where $\kappa$ is referred to as the \emph{spectral index} and $\vth$ is the \emph{thermal velocity}. In the limit $\kappa \rightarrow \infty$, the kappa distribution in \eqref{eq:kappa} reduces 
to a Maxwellian as in \eqref{eq:maxwell} with temperature $T$ given by
\begin{equation}
k_B T = \frac{m\vth^2}{2}.
\end{equation}

While the thermodynamic foundations leading to the Maxwell-Boltzmann distribution in \eqref{eq:maxwell} are well understood, this is not the case for the kappa distribution. The Maxwell-Boltzmann distribution 
follows from a straightforward application of Jaynes' maximum entropy principle~\cite{Jaynes1957}, where we search for the most unbiased distribution that agrees with known information, namely the distribution 
that maximizes the Boltzmann-Gibbs entropy
\begin{equation}
\label{eq:entropy}
\mathcal{S}[p] \defeq -k_B\int d\bm{\Gamma} p(\bm \Gamma)\ln p(\bm \Gamma)
\end{equation}
subject to the constraints that impose the information we have. Here $\bm \Gamma$ are the microstates of the system, commonly understood to be the full set of phase space coordinates 
$(\bm{r}_1, \ldots, \bm{r}_N, \bm{p}_1, \ldots, \bm{p}_N)$. However, this does not need to be the case: it is possible to apply this formalism to any subset of the degrees of freedom under any 
parameterization, and in particular it is possible to use the velocities $\bm{v}_i$ instead of the momenta $\bm{p}_i$. In fact, by taking $\bm \Gamma$ as the single-particle velocity $\bm{v}$, and 
maximizing $\mathcal{S}[p]$ for $p(\bm v)$ subject to the constraint of fixed single-particle kinetic energy
\begin{equation}
\left<\frac{m\bm{v}^2}{2}\right>_p = \int d\bm{v} p(\bm v) \left(\frac{m\bm{v}^2}{2}\right) = \overline{k}
\end{equation}
we readily obtain the Maxwell-Boltzmann distribution as
\begin{equation}
P(\bm v|\beta) = \frac{1}{Z(\beta)}\exp\left(-\frac{\beta m\bm{v}^2}{2}\right),
\end{equation}
which is \eqref{eq:maxwell} after normalization. An early theoretical attempt~\cite{Tsallis1988} to explain the origin of the kappa distribution and other so-called $q$-canonical distributions is known as 
Tsallis' non-extensive statistics~\cite{Tsallis2009}. This theory postulates a new functional that must be maximized under appropriate constraints instead of \eqref{eq:entropy}, the $q$-entropy, 
defined by
\begin{equation}
S_q[p] \defeq \frac{k_B}{q-1}\left(1-\int d\bm{\Gamma}\,p(\bm \Gamma)^q \right),
\end{equation}
where $q$ is known as the \emph{entropic index}. In the limit $q \rightarrow 1$, $S_q$ reduces to the Boltzmann-Gibbs entropy in \eqref{eq:entropy}.

However, the use of non-extensive statistics has not been free of controversy~\cite{Nauenberg2003, Presse2013, Presse2014, Tsallis2015, Jizba2019, Caticha2021, Marechal2024}. It has been argued~\cite{Nauenberg2003} 
that inconsistencies arise when placing in contact two systems with different values of $q$, and, moreover, that the use of generalized entropies (including but not limited to Tsallis entropy) introduces 
biases~\cite{Presse2013, Caticha2021} into statistical inference. On the topic of inference, that is, deriving a model by maximizing Tsallis entropy subject to constraints, there are also incompatibilities with the Bayesian rule 
that updates probabilities from prior to posterior~\cite{Presse2014}, as well as leading to non-convex optimization problems~\cite{Marechal2024}.

Fortunately, nowadays there are alternative formalisms that can explain the existence of kappa distributions by enhancing rather than rejecting the maximum entropy principle based on the Boltzmann-Gibbs entropy for 
inference. Among them, one of the most promising is the theory of superstatistics~\cite{Beck2003, Beck2004}, where the inverse temperature $\beta$ is promoted from a constant to a random variable having its own probability 
density. This mechanism brings about non-equilibrium ensembles, including the power laws such as \eqref{eq:kappa} commonly associated with non-extensive statistics, while also allowing for several other 
possibilities~\cite{Sanchez2021, Chung2020}. Superstatistics achieves this by considering mixtures of canonical ensembles, which are in turn obtained from maximization of the usual Boltzmann-Gibbs entropy~\cite{Pachter2024}.

In this work we show how the kappa distribution is recovered in the framework of superstatistics, but using an invariant parameterization of the distribution of inverse temperatures. This allows us to separate 
the effects of the temperature uncertainty from finite-size effects in the case of small number of particles. Some new results regarding the connection between correlations and the distance to equilibrium, 
the meaning of the superstatistical temperature and the entropy of kappa-distributed plasmas are also presented.

\newpage
\section{Superstatistics}

The formalism of superstatistics~\cite{Beck2003, Beck2004} is an alternative to non-extensive statistics that effectively and concisely recovers non-canonical ensembles, including but not limited to the $q$-canonical 
ensembles of non-extensive (Tsallis) statistics. It has been successfully applied to plasma physics~\cite{Ourabah2015, Davis2019b, Ourabah2020b, Sanchez2021, Gravanis2021}, condensed matter systems~\cite{Dixit2013, 
Dixit2015, Herron2021}, high-energy physics and gravitation~\cite{Jizba2010, Ayala2018, Ourabah2019}, among other fields~\cite{Chen2008, Denys2016, Bogachev2017, Schafer2018, Costa2022, Sanchez2025}.

In superstatistics, the inverse temperature $\beta$ is not a constant but a random variable with its own probability density. Here we will introduce the notation $P(A|\params)$ to indicate the superstatistical 
distribution of an arbitrary quantity $A$ given a collection of parameters denoted by $\params$, parameters that completely describe the superstatistical model. As an illustration, consider $\beta$ following a gamma 
distribution with shape parameter $k$ and scale parameter $\theta$. In this case we will write it as $P(\beta|k, \theta)$, hence $\params = (k, \theta)$. Using this notation, from here on a generic superstatistical 
probability density for $\beta$ will be written as $P(\beta|\params)$, while the superstatistical distribution of microstates (the superstatistical ``ensemble'') becomes $P(\bm \Gamma|\params)$ and the corresponding 
energy distribution is $P(E|\params)$. In particular, our invariant parameterization of the kappa distribution, originally introduced in Ref.~\cite{Davis2023e} and explained in the next section, uses $\params = (u, \beta_S)$ 
with $\beta_S$ the mean inverse temperature and $u$ its reduced variance.

\noindent
Superstatistics replaces the canonical ensemble that describes the distribution of microstates in equilibrium,
\begin{equation}
\label{eq:canonical}
P(\bm \Gamma|\beta) = \frac{\exp\big(-\beta\H(\bm \Gamma)\big)}{Z(\beta)},
\end{equation}
by a joint distribution $P(\bm \Gamma, \beta|\params)$ of the microstates $\bm \Gamma$ and the inverse temperature $\beta$, and this joint distribution is factorized as
\begin{equation}
\label{eq:joint}
P(\bm \Gamma, \beta|\params) = P(\beta|\params)P(\bm \Gamma|\beta, \params)
\end{equation}
using the product rule of probability theory. Note that this does not necessarily mean that $\beta$ is a fluctuating physical quantity of the system (i.e. a dynamical variable): in a more general, Bayesian 
formulation~\cite{Sattin2006}, the inverse temperature can be taken as an unknown parameter reflecting relevant information about the system. As we further discuss in Section~\ref{sec:observable}, it is in general 
impossible to define a phase-space function $B(\bm \Gamma)$ such that it corresponds one-to-one with the parameter $\beta$. Because $\params$ represents knowledge about $\beta$, we see that the state of knowledge 
$(\beta, \params)$ can be replaced by $\beta$ itself and thus we can replace $P(\bm \Gamma|\beta, \params)$ by $P(\bm \Gamma|\beta)$ as given by \eqref{eq:canonical}. Therefore, we can write \eqref{eq:joint} as
\begin{equation}
\label{eq:super_joint}
P(\bm \Gamma, \beta|\params) = P(\beta|\params)\frac{\exp\big(-\beta \H(\bm \Gamma)\big)}{Z(\beta)},
\end{equation}
and the superstatistical distribution of microstates $P(\bm \Gamma|\params)$ as the integral over $\beta$ of \eqref{eq:super_joint}, namely
\begin{equation}
\label{eq:super_ensemble}
P(\bm \Gamma|\params) = \int_0^\infty d\beta\,P(\beta|\params)\frac{\exp\big(-\beta \H(\bm \Gamma)\big)}{Z(\beta)}.
\end{equation}

The integration over $\beta$ here is conceptually just the marginalization of a \emph{nuisance parameter} in the Bayesian sense, which is different from the treatment in a dynamical interpretation of superstatistics, 
as originally presented by Beck and Cohen, and as discussed by Abe~\cite{Abe2014b}. This means \eqref{eq:super_ensemble} is simply a consequence of the sum and product rule of probability theory, and therefore is always 
valid, provided that the canonical ensemble is well-defined for the Hamiltonian $\H$.

From \eqref{eq:super_ensemble} we see that superstatistical ensembles belong to a class of non-equilibrium steady-state ensembles, where the microstate distribution depends only on the Hamiltonian, that is, where
\begin{equation}
\label{eq:rho}
P(\bm \Gamma|\params) = \rho\big(\H(\bm \Gamma); \params\big),
\end{equation}
with $\rho(E; \params)$ the \emph{ensemble function} $\rho(E; \params)$. By defining the \emph{superstatistical weight function}
\begin{equation}
f(\beta; \params) \defeq \frac{P(\beta|\params)}{Z(\beta)},
\end{equation}
it is clear that $\rho$ is the Laplace transform of $f$,
\begin{equation}
\rho(E; \params) = \int_0^\infty d\beta\,f(\beta; \params)\exp(-\beta E).
\end{equation}

The canonical ensemble is a particular case of superstatistics where $\beta$ becomes again a constant, that is, when the uncertainty over $\beta$ vanishes. This is of course achieved for
\begin{equation}
\label{eq:prob_beta_delta}
P(\beta|\beta_0) = \delta(\beta-\beta_0),
\end{equation}
as in this case $P(\bm \Gamma|\params)$ in \eqref{eq:super_ensemble} reduces to \eqref{eq:canonical} with $\beta = \beta_0$. In general, because $\beta$ and $\bm \Gamma$ are correlated in 
\eqref{eq:super_joint}, the marginal distribution $P(\bm \Gamma|\params)$ is different from the canonical ensemble. Superstatistics then produces non-canonical ensembles 
by postulating different weight functions, or equivalently, different inverse temperature distributions.

We will now define two quantities of interest for superstatistical steady states. First, the mean inverse temperature of the ensemble, denoted by $\beta_S$, will be the mean value of 
$\beta$ under the distribution $P(\beta|\params)$, that is,
\begin{equation}
\label{eq:betaS}
\beta_S \defeq \big<\beta\big>_{\params} = \int_0^\infty d\beta\,P(\beta|\params)\beta.
\end{equation}

\noindent
Secondly, we will define the variance of $\beta$ under the distribution $P(\beta|\params)$ as
\begin{equation}
\label{eq:usuper}
\U \defeq \big<(\delta \beta)^2\big>_{\params} = \int_0^\infty d\beta\,P(\beta|\params)\big(\beta-\beta_S\big)^2,
\end{equation}
noting that $\U$ as a variance in superstatistics is a particular case of a more general definition, namely the covariance
\begin{equation}
\U \defeq \big<\delta \beta_F\,\delta \beta_\Omega\big>_{\params} = \big<\beta_F\,\beta_\Omega\big>_{\params} - (\beta_S)^2
\end{equation}
between two different but complementary definitions of inverse temperature, the \emph{fundamental inverse temperature}~\cite{Davis2023b}
\begin{equation}
\beta_F(E; \params) \defeq -\frac{\partial}{\partial E}\ln \rho(E; \params)
\end{equation}
with $\rho(E; \params)$ defined by \eqref{eq:rho}, and the \emph{microcanonical inverse temperature}
\begin{equation}
\beta_\Omega(E) \defeq \frac{\partial}{\partial E}\ln \Omega(E),
\end{equation}
with $\Omega(E) = \int d\bm{\Gamma}\,\delta\big(E-\H(\bm \Gamma)\big)$ the density of states. This inverse temperature covariance $\U$ is non-negative for superstatistics, as is clear from \eqref{eq:usuper}, but can be negative 
for non-equilibrium steady-state models outside of superstatistics~\cite{Davis2022, Davis2022b}. Using $\U$ we can define the relative variance $u$ as
\begin{equation}
\label{eq:u}
u \defeq \frac{\U}{(\beta_S)^2},
\end{equation}
such that
\begin{equation}
0 \leq u \leq \frac{1}{2}
\end{equation}
for kappa-distributed particles. As $u$ measures the variance of the inverse temperature, higher values of $u$ indicate additional uncertainty regarding the energies of a particle system, with respect to the uncertainty 
already present in the Maxwellian distribution. On the other hand, higher values of $u$ also indicate a stronger departure from equilibrium.

The distribution $P(\beta|\params)$ has been shown~\cite{Davis2022b} to be invariant under the choice of subsystems in a composite system, and in particular it is independent of the number of particles in 
a multi-particle system. Nevertheless, it is possible to obtain a size-dependent ensemble function $\rho_N(E; \params)$ as the Laplace transform of the size-dependent weight function,
\begin{equation}
f_N(\beta; \params) = \frac{P(\beta|\params)}{Z_N(\beta)},
\end{equation}
given a size-dependent partition function $Z_N(\beta)$. The invariance of $P(\beta|\params)$ means that its moments $\beta_S$ and $\U$ are promising candidates for any parameterization of a superstatistical 
state. The same is true of $u$, being a combination of $\U$ and $\beta_S$. In particular, using $u$ as one of the superstatistical parameters is useful because the limit $u \rightarrow 0$ turns 
$P(\beta|\params)$ into a Dirac delta distribution, therefore being the canonical (equilibrium) limit of superstatistics.

\section{Kappa distribution as a superstatistical mixture of Maxwellians}

Consider a two-species plasma consisting of $N = N_A + N_B$ particles, where $N_A$ and $N_B$ are the number of particles of species $A$ and $B$, respectively, and where $m_A$ and $m_B$ are the corresponding 
masses for each of the species. That is, the mass $m_i$ of the $i$-th particle with $i=1,2,\ldots,N$ is such that $m_i \in \{m_A, m_B\}$. The $3N$-dimensional vector of all particle velocities will be denoted by 
$\bm V \defeq (\bm{v}_1, \bm{v}_2, \ldots, \bm{v}_N)$, with the probability density of $\bm V$ being the $N$-particle Maxwell-Boltzmann distribution,
\begin{equation}
\label{eq:multi_maxwell}
P(\bm V|\beta) = \prod_{i=1}^N P(\bm{v}_i|\beta, m_i) = \left(\frac{\tilde{m}\beta}{2\pi}\right)^{\frac{3N}{2}}\exp\big(-\beta K(\bm V)\big),
\end{equation}
which is just the product of $N$ independent single-particle distributions $P(\bm v|\beta, m_i)$ as in \eqref{eq:maxwell}. Here
\begin{equation}
\label{eq:geom_mass}
\tilde{m} \defeq \left(\prod_{i=1}^N m_i\right)^{\frac{1}{N}} = m_A^{N_A/N}\cdot m_B^{N_B/N},
\end{equation}
is the geometric mean of the particle masses and
\begin{equation}
K(\bm V) \defeq \sum_{i=1}^N \frac{m_i \bm{v}_i^2}{2}
\end{equation}
is the kinetic energy of the $N$-particle system. In order to move from the Maxwellian to the kappa distribution within the superstatistical framework, we need to postulate a distribution $P(\beta|\params)$ 
of inverse temperatures. The particular distribution that achieves this is a gamma distribution, which following Ref.~\cite{Davis2023e} we will write as
\begin{equation}
\label{eq:prob_beta_kappa}
P(\beta|u, \beta_S) = \frac{1}{u\beta_S\Gamma\left(\frac{1}{u}\right)}\exp\left(-\frac{\beta}{u\beta_S}\right)\left(\frac{\beta}{u\beta_S}\right)^{\frac{1}{u}-1}.
\end{equation}
 
This fixes our superstatistical parameters to be $\params = (u, \beta_S)$, with $u$ and $\beta_S$ as defined in \eqref{eq:u} and \eqref{eq:betaS}, respectively. The most probable value of $\beta$ under 
this distribution is
\begin{equation}
\label{eq:beta_mode}
\beta^* \defeq \beta_S\,(1-u),
\end{equation}
while the mean and variance of $\beta$ are given by
\begin{subequations}
\begin{align}
\label{eq:beta_mean}
\big<\beta\big>_{u, \beta_S} & = \beta_S, \\
\big<(\delta \beta)^2\big>_{u, \beta_S} & = u (\beta_S)^2,
\end{align}
\end{subequations}
as expected. If we replace \eqref{eq:prob_beta_kappa} into \eqref{eq:super_ensemble} and introduce the variable
\begin{equation}
t \defeq \frac{\beta}{u\beta_S},
\end{equation}
we obtain
\begin{equation}
\label{eq:prob_V_pre}
P(\bm V|u, \beta_S) = \frac{(u\beta_S)^{\frac{3N}{2}}}{\Gamma\left(\frac{1}{u}\right)}\left(\frac{\tilde{m}}{2\pi}\right)^{\frac{3N}{2}}
\int_0^\infty dt\,t^{\frac{1}{u}+\frac{3N}{2}-1}\exp\left(-t\Big[1+ u\beta_S K(\bm V)\Big]\right),
\end{equation}
and after performing the integral in $t$ and recognizing the canonical partition function
\begin{equation}
Z_N(\beta) = \int d\bm{V}\,\exp\big(-\beta K(\bm V)\big) = \left(\frac{2\pi}{\tilde{m}\beta}\right)^{\frac{3N}{2}},
\end{equation}
we recover the multi-particle kappa distribution~\cite{Livadiotis2019} in the form
\begin{equation}
\label{eq:multi_kappa}
P(\bm V|u, \beta_S) = \frac{C_N(u)}{Z_N(\beta_S)}\Big[1 + u\beta_S K(\bm V)\Big]^{-\left(\frac{1}{u}+\frac{3N}{2}\right)}.
\end{equation}
where we have defined the constant
\begin{equation}
C_N(u) \defeq \frac{u^{\frac{3N}{2}} \Gamma\left(\frac{3N}{2}+\frac{1}{u}\right)}{\Gamma\left(\frac{1}{u}\right)},
\end{equation}
for convenience. Comparison of the negative exponent in \eqref{eq:multi_kappa} with the kappa distribution yields a size-dependent spectral index
\begin{equation}
\kappa_N \defeq \frac{1}{u} + \frac{3N}{2}-1.
\end{equation}

Here it is clear that the dependence on $N$ of $\kappa_N$ is due to the partition function $Z_N$ and not the distribution $P(\beta|u, \beta_S)$. Also note that, from the Stirling approximation,
\begin{equation}
\lim_{u \rightarrow 0} C_N(u) = 1
\end{equation}
for all $N$, and also
\begin{equation}
\lim_{u \rightarrow 0} \Big[1+u\beta_S K\Big]^{-\left(\frac{1}{u}+\frac{3N}{2}\right)} = \exp(-\beta_S K),
\end{equation}
thus confirming that the limit $u \rightarrow 0$ of \eqref{eq:multi_kappa} yields the Maxwell-Boltzmann distribution in \eqref{eq:multi_maxwell} with 
$\beta = \beta_S$. For a single particle, \eqref{eq:multi_kappa} reduces to
\begin{equation}
\label{eq:kappa_new}
P(\bm v|u, \beta_S) = \left(\frac{m u \beta_S}{2\pi}\right)^{\frac{3}{2}} \frac{\Gamma\left(\frac{3}{2}+\frac{1}{u}\right)}{\Gamma\left(\frac{1}{u}\right)}
\left[1 + u\beta_S \frac{m\bm{v}^2}{2}\right]^{-\left(\frac{1}{u}+\frac{3}{2}\right)}
\end{equation}
which is exactly the kappa distribution as defined in \eqref{eq:kappa}, under the substitutions
\begin{equation}
\kappa \defeq \frac{1}{u} + \frac{1}{2} = \kappa_1,
\end{equation}
and
\begin{equation}
\frac{m\vth^2}{2} \defeq \frac{1}{\beta^*}.
\end{equation}

\section{Moments of the kappa distribution of energies from superstatistics}

One of the advantages of the superstatistical framework is that in many cases it simplifies the computation of expectation values. For instance, we can easily obtain all the moments of 
the $N$-particle energy distribution associated to \eqref{eq:multi_kappa} without computing the distribution explicitly, by using the moments of the canonical energy distribution
\begin{equation}
P(K|\beta) = \int d\bm{V} P(\bm V|\beta)\delta\big(K(\bm V)-K\big) = \frac{\beta^{\frac{3N}{2}}}{\Gamma\left(\frac{3N}{2}\right)} \exp(-\beta K)K^{\frac{3N}{2}-1},
\end{equation}
which is a gamma distribution. Recalling that for a gamma distribution of the form
\begin{equation}
P(X|r, \theta) = \frac{\exp(-X/\theta)X^{r-1}}{\Gamma(r)\theta^r}
\end{equation}
the $n$-th moment is given by
\begin{equation}
\label{eq:gamma_moments}
\big<X^n\big>_{r, \theta} = \frac{1}{\Gamma(r)\theta^r}\int_0^\infty dX \exp(-X/\theta)X^{r+n-1} = \theta^n\,\frac{\Gamma(r+n)}{\Gamma(r)}, \qquad n > -r,
\end{equation}
we obtain the canonical energy moments as
\begin{equation}
\label{eq:moments_canon}
\big<K^n\big>_\beta = \beta^{-n}\,\frac{\Gamma\left(\frac{3N}{2}+n\right)}{\Gamma\left(\frac{3N}{2}\right)}, \qquad n > -\frac{3N}{2}.
\end{equation}

\noindent
Taking expectation over $P(\beta|u, \beta_S)$ we have
\begin{equation}
\label{eq:moments_pre}
\big<K^n\big>_{u, \beta_S} = \big<\beta^{-n}\big>_{u, \beta_S}\,\frac{\Gamma\left(\frac{3N}{2}+n\right)}{\Gamma\left(\frac{3N}{2}\right)}
\end{equation}
and by using the fact that $P(\beta|u, \beta_S)$ in \eqref{eq:prob_beta_kappa} is also a gamma distribution, we use \eqref{eq:gamma_moments} to easily get
\begin{equation}
\label{eq:moments_beta}
\big<\beta^n\big>_{u, \beta_S} = (u\beta_S)^n \;\frac{\Gamma\left(\frac{1}{u}+n\right)}{\Gamma\left(\frac{1}{u}\right)}.
\end{equation}

\noindent
Replacing \eqref{eq:moments_beta} into \eqref{eq:moments_pre} we obtain the moments of $P(K|u, \beta_S)$ as
\begin{equation}
\label{eq:moments_K}
\big<K^n\big>_{u, \beta_S} = (u\beta_S)^{-n}\,\frac{\Gamma\left(\frac{3N}{2}+n\right)}{\Gamma\left(\frac{3N}{2}\right)}\frac{\Gamma\left(\frac{1}{u}-n\right)}{\Gamma\left(\frac{1}{u}\right)}, \qquad -\frac{3N}{2} < n < \frac{1}{u}.
\end{equation}

The upper bound for $n$ imposes an upper bound for $u$ (lower bound for $\kappa$). For instance, in order for the kinetic energy to have a well-defined variance, we must have $u < 1/2$ 
(equivalently, $\kappa > 5/2$). On the other hand, requiring the existence of higher moments of $K$ imposes an even more strict bound on $u$. Replacing $n = 1$ in \eqref{eq:moments_K} and using the recursive property
of the gamma function,
\begin{equation}
z\Gamma(z) = \Gamma(z+1),
\end{equation}
we obtain the mean kinetic energy of the system,
\begin{equation}
\label{eq:mean_K}
K_S \defeq \big<K\big>_{u, \beta_S} = \frac{3N}{2\beta_S (1-u)} = \frac{3N}{2\beta^*},
\end{equation}
and we see that the $N$-particle kappa distribution follows a form of equipartition theorem where the equipartition temperature is defined by
\begin{equation}
k_B T_{\text{equip}} \defeq \frac{k_B T_S}{1-u}.
\end{equation}

\noindent
Similarly, using \eqref{eq:moments_K} with $n = 2$ and some algebra we obtain the relative variance of $K$,
\begin{equation}
\frac{\big<(\delta K)^2\big>_{u, \beta_S}}{(K_S)^2} = \frac{(3N-2)u + 2}{3N(1-2u)}
\end{equation}
which does not vanish in the thermodynamic limit unless $u = 0$, due to the uncertainty in $\beta$. For $N = 1$, the mean and variance of the single-particle kinetic energy are given by
\begin{subequations}
\begin{align}
k_S & \defeq \big<k\big>_{u, \beta_S} = \frac{3}{2 \beta^*}, \\
\sigma_k^2 & \defeq \big<(\delta k)^2\big>_{u, \beta_S} = \frac{u+2}{3(1-2u)} (k_S)^2,
\end{align}
\end{subequations}
respectively, thus from $k_S$ and $\sigma_k^2$ we can determine $u$ and $\beta_S$ as
\begin{subequations}
\begin{align}
u & = \frac{3\sigma_k^2 - 2 k_S^2}{6\sigma_k^2 + k_S^2}, \\
\beta_S & = \frac{3}{2(1-u)k_S}.
\end{align}
\end{subequations}

\noindent
On the other hand, for $u < 1/3$, that is, $\kappa > 7/2$, we can also define the single-particle kinetic energy \emph{skewness}, given by
\begin{equation}
\text{Skew}(u) \defeq \frac{1}{\sigma_k^3} \left<\left(\frac{m\bm{v}^2}{2} - k_S\right)^3\right>_{u, \beta_S} = \frac{4(1+2u)}{1-3u}\sqrt{\frac{1-2u}{3(2+u)}}.
\end{equation}

\begin{figure}[t!]
\begin{center}
\includegraphics[width=0.49\textwidth]{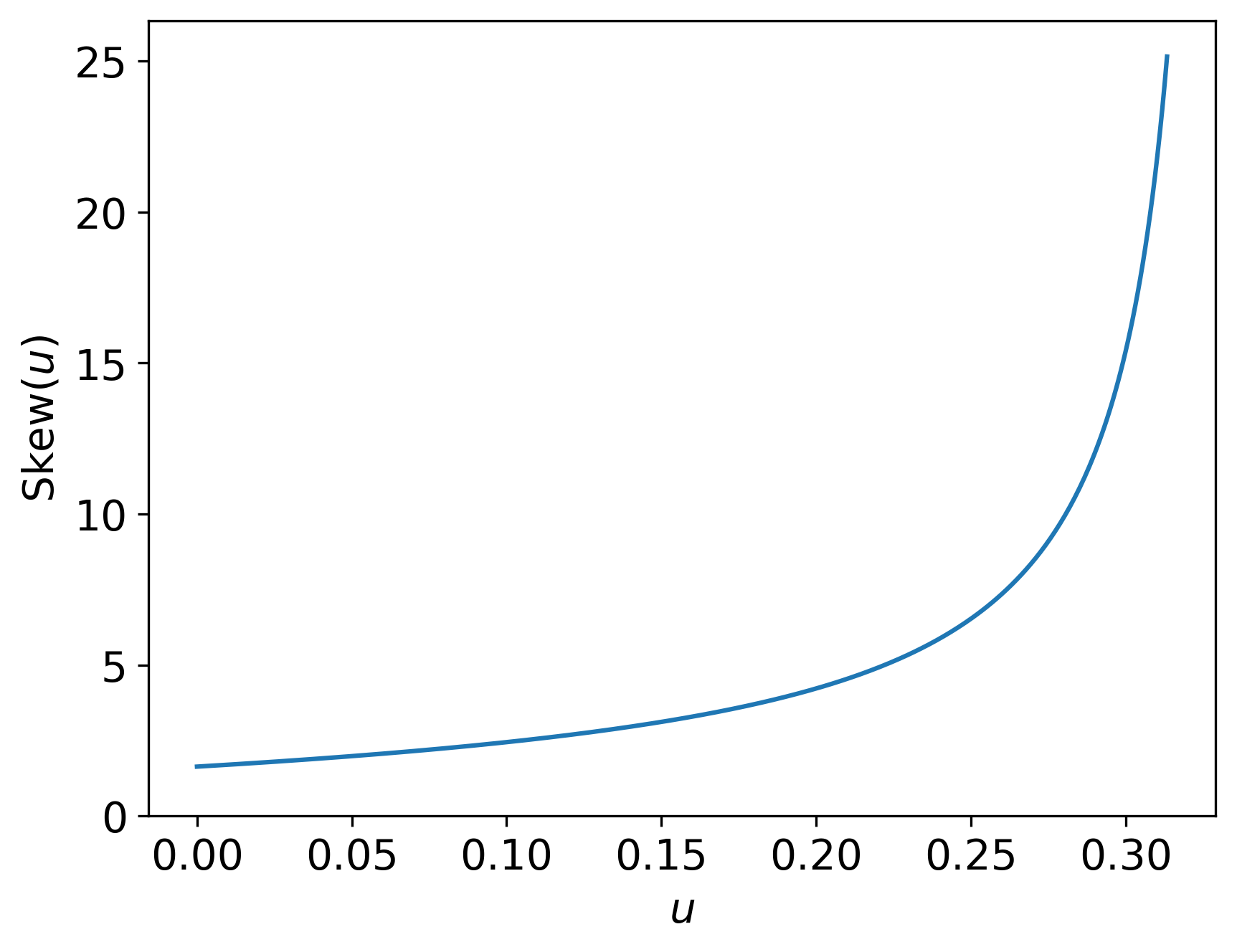}
\includegraphics[width=0.49\textwidth]{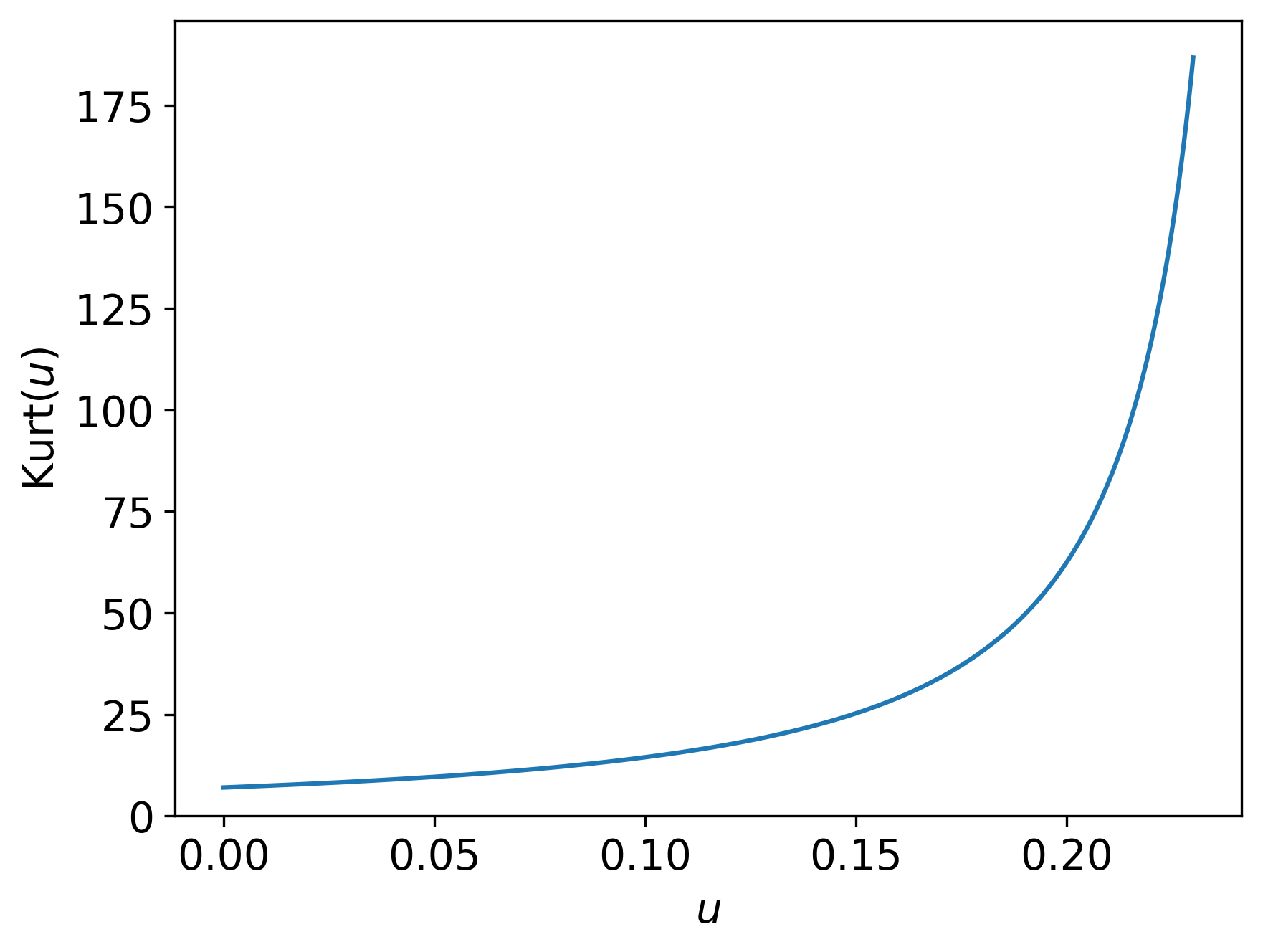}
\end{center}
\caption{Left, skewness of the kinetic energy distribution as a function of $u$. Right, kurtosis of the same distribution of kinetic energies.}
\label{fig:skewkurt}
\end{figure}

We note that the skewness only depends on the value of $u$. It is also non-negative, in accordance with the fact that the kinetic energy tails always extend to the right, and increases monotonically and without bound with 
$u$ from its minimum value $\text{Skew}(0) = 2\sqrt{2/3}$ at $u = 0$ for $u < 1/3$, as shown in the left panel of Fig.~\ref{fig:skewkurt}. Similarly, the kurtosis of the distribution of kinetic energies is also a 
monotonically increasing function of $u$, given by
\begin{equation}
\text{Kurt}(u) \defeq \frac{1}{\sigma_k^4} \left<\left(\frac{m\bm{v}^2}{2} - k_S\right)^4\right>_{u, \beta_S} = \frac{(1-2u)\big[14+u(17+23u)\big]}{(u+2)(1-3u)(1-4u)},
\end{equation}
as shown in the right panel of Fig.~\ref{fig:skewkurt}. As usual, higher values of kurtosis indicate heavier kinetic energy tails when compared with the Maxwellian distribution.
 
\section{Correlation between particle kinetic energies}

The kinetic energies of particles in a multi-particle kappa distribution such as \eqref{eq:multi_kappa} are correlated due to the presence of inverse temperature variations, and in 
fact, the existence of a linear dependence between kinetic energies has been shown to be a sufficient condition to obtain the kappa distribution~\cite{Davis2023e}. 

In this section we will evaluate this correlation between the energies of two trial particles with energies $k_1$ and $k_2$, through the covariance
\begin{equation}
\big<\delta k_1 \delta k_2\big>_{u, \beta_S} \defeq \big<k_1 k_2\big>_{u, \beta_S} - \big<k_1\big>_{u, \beta_S} \big<k_2\big>_{u, \beta_S}.
\end{equation}

\noindent
First we use the fact that kinetic energies in the canonical ensemble are uncorrelated, therefore $\big<\delta k_1\delta k_2\big>_\beta = 0$ and
\begin{equation}
\big<k_1 k_2\big>_\beta = \big<k_1\big>_\beta \; \big<k_2\big>_\beta.
\end{equation}
Now, because the single-particle kinetic energies are identically distributed,
\begin{equation}
\big<k_1\big>_\beta = \big<k_2\big>_\beta = \big<k\big>_\beta = \frac{3}{2\beta}
\end{equation}
and we have
\begin{equation}
\big<k_1 k_2\big>_\beta = \left(\frac{3}{2\beta}\right)^2,
\end{equation}
hence taking expectation under $P(\beta|u, \beta_S)$ we have
\begin{equation}
\big<k_1 k_2\big>_{u, \beta_S} = \left< \big<k_1 k_2\big>_\beta\right>_{u, \beta_S} = \left(\frac{3}{2}\right)^2 \big<\beta^{-2}\big>_{u, \beta_S}
= \left(\frac{3}{2\beta_S}\right)^2\;\frac{1}{(1-u)(1-2u)}
\end{equation}
where we have used \eqref{eq:moments_beta} for $n = -2$. Therefore the covariance between $k_1$ and $k_2$ is
\begin{equation}
\label{eq:k1_k2_cov}
\big<\delta k_1 \delta k_2\big>_{u, \beta_S} = \frac{u}{1-2u}(k_S)^2.
\end{equation}

We see that the covariance is positive for $0 < u < 1/2$ and moreover, increases strictly with $u$, being zero only for $u = 0$. This reflects a well-known connection~\cite{Nicolaou2020c, Livadiotis2022, Davis2023e} 
between interparticle correlations and the presence of kappa distributions, and in particular that correlations increase with the distance from Maxwellian equilibrium as measured by $\kappa$ (or equivalently, $u$). 
From \eqref{eq:k1_k2_cov} we can compute the Pearson correlation, which yields
\begin{equation}
\label{eq:pearson}
\rho_{k_1, k_2}(u) \defeq \frac{\big<\delta k_1 \delta k_2\big>_{u, \beta_S}}{\big<(\delta k)^2\big>_{u, \beta_S}} = \frac{3u}{u+2} = \frac{3}{2\kappa},
\end{equation}
and also strictly increases with $u$ (decreases with $\kappa$). However, a more robust indicator of correlation is the mutual information $I_{k_1, k_2}$ between $k_1$ and $k_2$. 
The mutual information~\cite{CoverThomas2006} between two variables $X$ and $Y$ is defined by
\begin{equation}
I(X; Y) \defeq \left<\ln \frac{P(X, Y|I)}{P(X|I)P(Y|I)}\right>_I = \int dX dY P(X, Y|I)\ln \frac{P(X, Y|I)}{P(X|I)P(Y|I)},
\end{equation}
and is equal to the Kullback-Leibler distance between the joint distribution of $X$ and $Y$ and the model that assumes statistical independence between $X$ and $Y$, namely the product of the marginal distributions. 
As such, $I(X; Y)$ measures the amount of statistical dependence or correlation between the variables, being such that $I(X; Y) \geq 0$ with equality only when $X$ and $Y$ are independent. Alternatively, $I(X; Y)$ also 
represents the amount of information that one variable possesses about the other. In the case of kinetic energies $k_1$ and $k_2$ for kappa-distributed velocities, the mutual information takes the form
\begin{equation}
\label{eq:mutinf}
I_{k_1, k_2}(u, \beta_S) \defeq \left<\ln \left[\frac{P(k_1, k_2|u, \beta_S)}{P(k_1|u, \beta_S) P(k_2|u, \beta_S)}\right]\right>_{u, \beta_S}.
\end{equation}

\begin{figure}[t!]
\begin{center}
\includegraphics[width=0.5\textwidth]{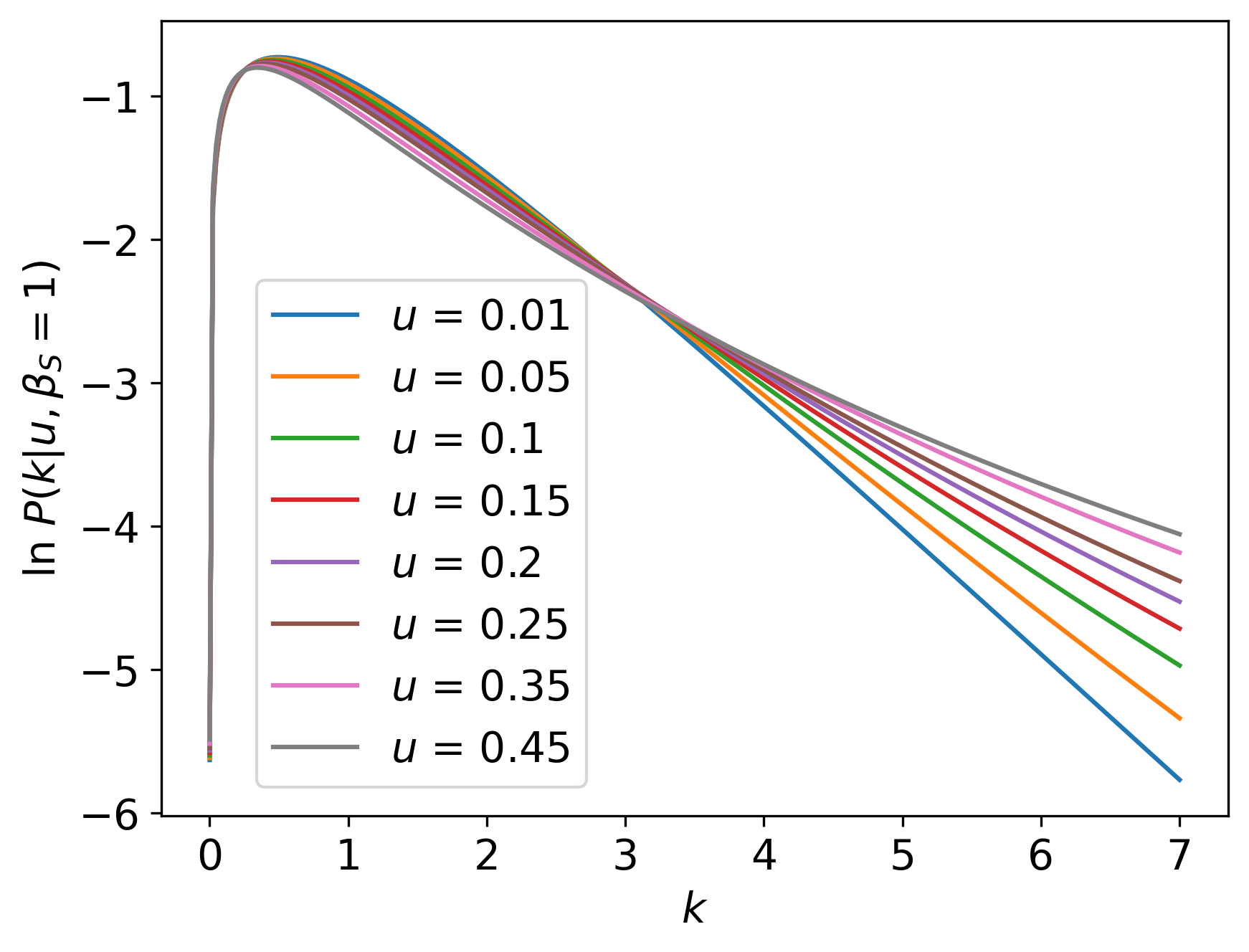}
\end{center}
\caption{Kinetic energy distribution for $\beta_S = 1$ and different values of $u$ between 0 and 1/2, according to \eqref{eq:marg_k1}.}
\label{fig:kindist}
\end{figure}

This quantifies the shared information between the random variables $k_1$ and $k_2$. From \eqref{eq:kappa_new} we readily obtain the distribution of single-particle kinetic energies
\begin{equation}
\label{eq:marg_k1}
P(k|u, \beta_S) = \int d\bm{v}P(\bm v|u, \beta_S)\delta\left(k-\frac{m\bm{v}^2}{2}\right) = \frac{2(u\beta_S)^{\frac{3}{2}}\,\Gamma\left(\frac{3}{2}+\frac{1}{u}\right)}{\sqrt{\pi}\,\Gamma\left(\frac{1}{u}\right)}\Big[1 + u\beta_S k\Big]^{-\left(\frac{1}{u}+\frac{3}{2}\right)} \sqrt{k},
\end{equation}
which is shown in Fig.~\ref{fig:kindist}. Interestingly, this distribution has an inflection point at a value of kinetic energy $k = k_c$ given by
\begin{equation}
k_c(u, \beta_S) \defeq \frac{1}{\beta_S\left(\sqrt{u(2+3u)}-u\right)}
\end{equation}
such that for $k < k_c$ the distribution is concave (consistent with a maximum) while it becomes convex for $k > k_c$. In contrast with the Maxwellian distribution of energies, which is concave everywhere, the value $k_c$ 
may be used to define a point beyond which the heavy tails of the distribution become relevant. 

\noindent
The joint distribution of $k_1$ and $k_2$ can be computed from \eqref{eq:multi_kappa} as
\begin{equation}
\label{eq:joint_k1_k2}
\begin{split}
P(k_1, k_2|u, \beta_S) & = \int d\bm{v}_1 d\bm{v}_2 P(\bm{v}_1, \bm{v}_2|u, \beta_S)\delta\left(k_1-\frac{m_1\bm{v}_1^2}{2}\right)\delta\left(k_2-\frac{m_2\bm{v}_2^2}{2}\right) \\
& = \frac{4(\beta_S)^3\,(1+u)(1+2u)}{\pi} \Big[1+u\beta_S(k_1+k_2)\Big]^{-\frac{1}{u}-3} \sqrt{k_1\; k_2}.
\end{split}
\end{equation}

\noindent
Upon replacing \eqref{eq:joint_k1_k2} and \eqref{eq:marg_k1} into \eqref{eq:mutinf}, we can finally evaluate $I_{k_1, k_2}$ as
\begin{equation}
\label{eq:mutual}
I_{k_1, k_2}(u) = 2F\left(\frac{1}{u}+\frac{3}{2}\right) - F\left(\frac{1}{u}\right) - F\left(\frac{1}{u}+3\right),
\end{equation}
where we have defined
\begin{equation}
F(z) \defeq z\psi(z) - \ln\,\Gamma(z)
\end{equation}
with $\psi$ the digamma function,
\begin{equation}
\psi(z) \defeq \frac{d}{dz}\ln \Gamma(z).
\end{equation}

\noindent
Notice that $I_{k_1, k_2}(u)$ can also be written as
\begin{equation}
I_{k_1, k_2}(u) = 2\D(u; 1)-D(u; 2),
\end{equation}
where $\D(u; N)$ is the superstatistical distance to equilibrium~\cite{Davis2024}, which in the case of the kappa distribution becomes
\begin{equation}
\D(u; N) \defeq \left<-\ln \left[\frac{P(\bm V|u, \beta_S)}{P(\bm V|\beta)}\right]\right>_{u, \beta_S} = F\left(\frac{1}{u}+\frac{3N}{2}\right) - F\left(\frac{1}{u}\right) - \frac{3N}{2}.
\end{equation}

\begin{figure}[t!]
\begin{center}
\includegraphics[width=0.475\textwidth]{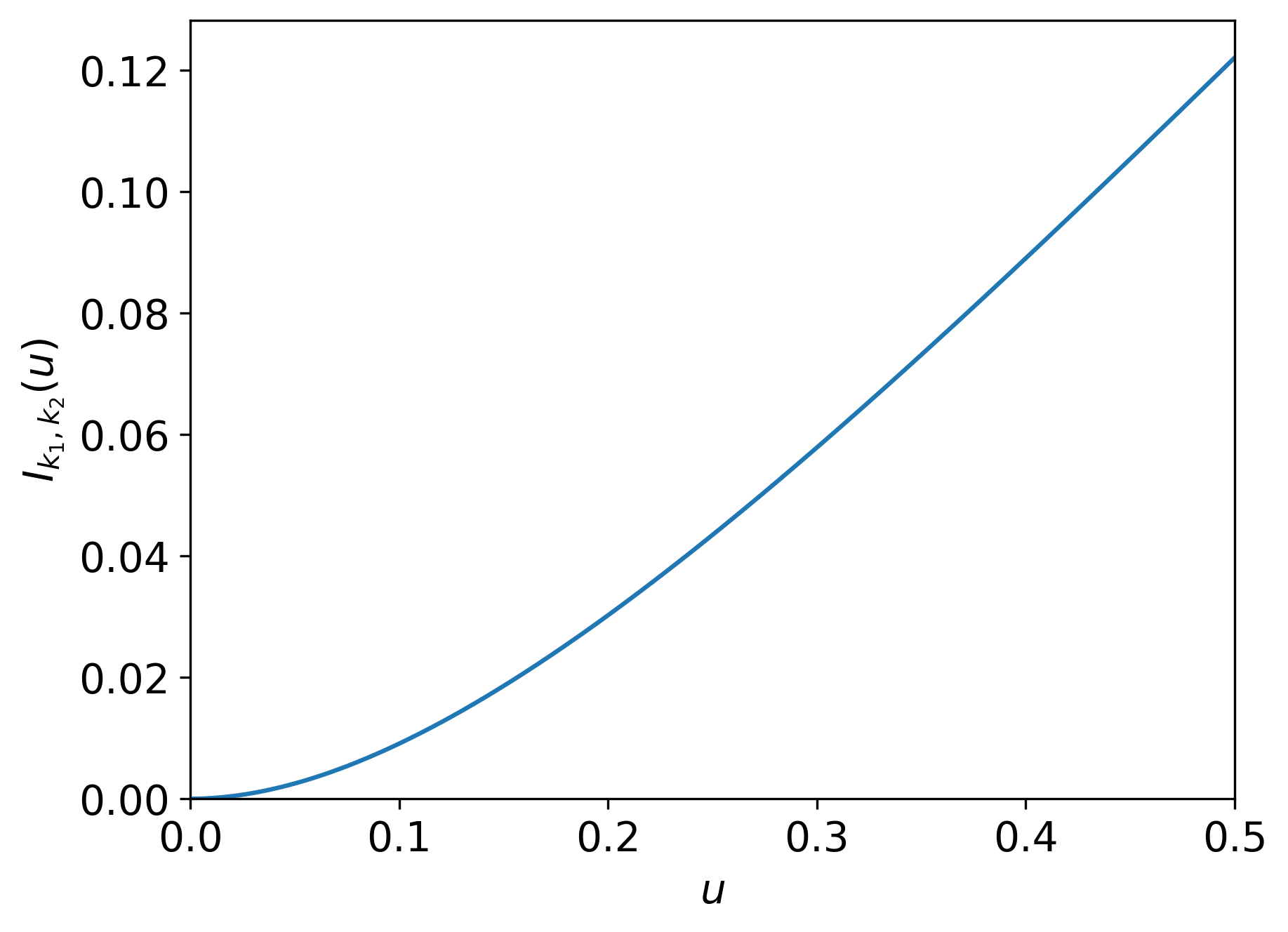}
\includegraphics[width=0.475\textwidth]{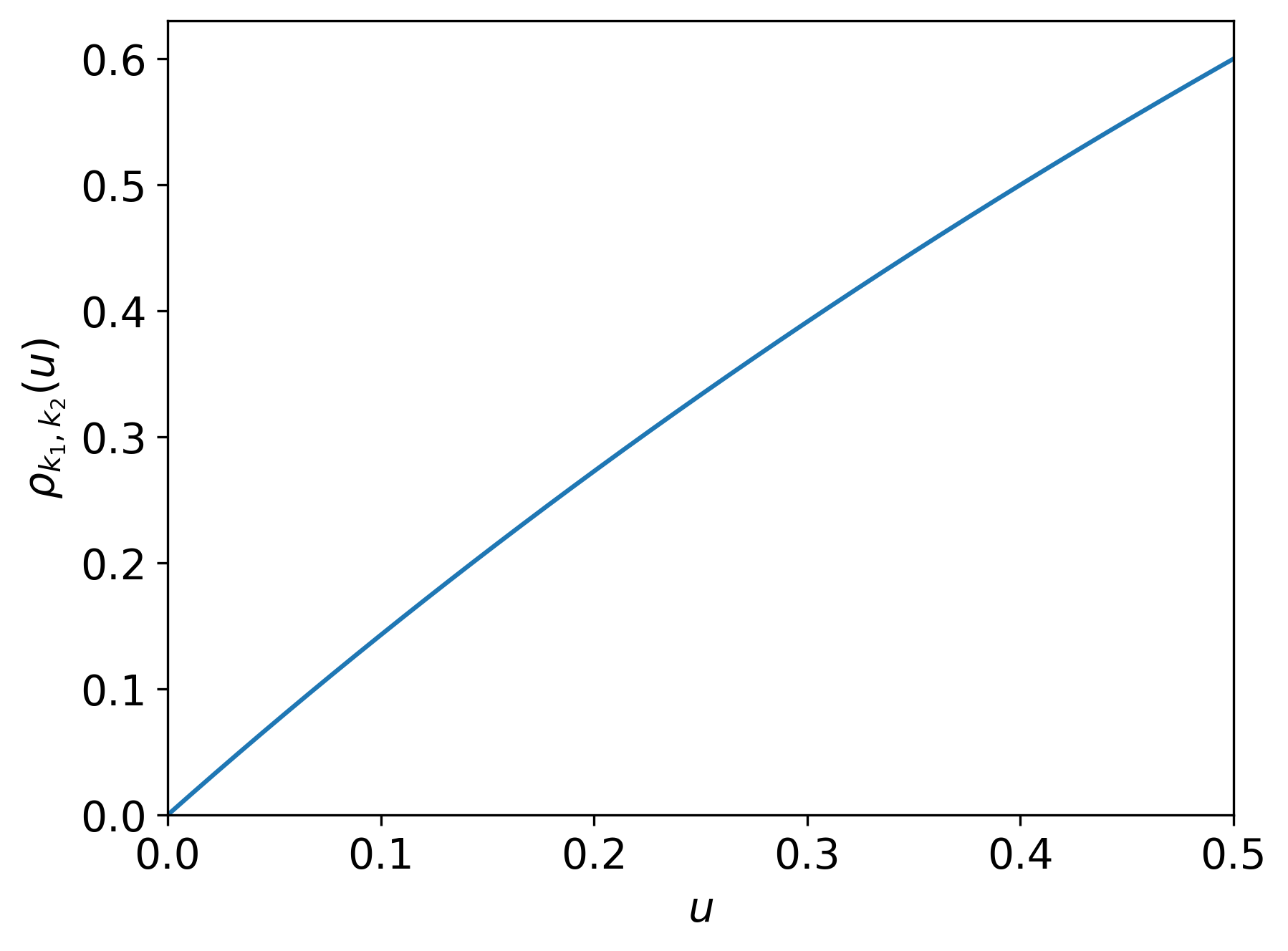}
\end{center}
\caption{Left, mutual information $I_{k_1, k_2}(u)$ in \eqref{eq:mutual} as a function of $u$. Right, Pearson correlation $\rho_{k_1, k_2}(u)$ in \eqref{eq:pearson} as a function of $u$.}
\label{fig:mutual_pearson}
\end{figure}

Because $F'(z) = z\psi'(z) > 0$ for $z > 0$ we can show that $I_{k_1, k_2}(u)$ strictly increases with $u$, and by combining the asymptotic approximations~\cite{Abramowitz1972} for $\psi(z)$ 
and $\ln \Gamma(z)$, namely
\begin{subequations}
\begin{align}
\psi(z) & \approx \ln z - \frac{1}{2z}, \\
\ln \Gamma(z) & \approx z\ln z - z + \frac{1}{2}\ln\,\left(\frac{2\pi}{z}\right),
\end{align}
\end{subequations}
for large values of $z$, we have that
\begin{equation}
F(z) \approx z + \frac{1}{2}\ln z - \frac{1}{2}\Big(\ln\,(2\pi) + 1\Big) 
\end{equation}
and we can verify that
\begin{equation}
\lim_{u \rightarrow 0} I_{k_1, k_2}(u) = 0,
\end{equation}
thus confirming that $k_1$ and $k_2$ are uncorrelated only in the canonical limit. Fig.~\ref{fig:mutual_pearson} shows $I_{k_1, k_2}$ and $\rho_{k_1, k_2}$ as a function of $u$.

\section{Observable temperature for kappa distributions}
\label{sec:observable}

It has been established~\cite{Davis2018} as a general property of non-canonical superstatistics that no function $\hat{\beta}(\bm \Gamma)$ exists being interchangeable with $\beta$ in 
the sense that
\begin{equation}
\big<G\big(\hat{\beta}(\bm \Gamma)\big)\big>_{\params} = \big<G(\beta)\big>_{\params}
\end{equation}
for an arbitrary function $G$. A consequence of this theorem is that $P(\beta|\params)$ cannot be written as a sampling distribution, in other words, there is no function 
$\hat{\beta}(\bm \Gamma)$ such that
\begin{equation}
\label{eq:sampling}
P(\beta|\params) = \big<\delta\big(\hat{\beta}(\bm \Gamma)-\beta\big)\big>_{\params}.
\end{equation}
 
One way to understand this impossibility theorem in the case of kappa distributions, is to notice that the uncertainty in $\beta$ does not vanish even when knowing the velocities 
$\bm V$ of all the particles. In fact, if such a function $\hat{\beta}(\bm V)$ existed for which \eqref{eq:sampling} with $\bm \Gamma = \bm V$ and $\params = (\bm V, u, \beta_S)$ 
would hold, it would imply
\begin{equation}
\label{eq:beta_no_uncertainty}
P(\beta|\bm V, u, \beta_S) = \delta\big(\beta - \hat{\beta}(\bm V)\big).
\end{equation}

In the case of our multi-particle kappa distribution, following Ref.~\cite{Davis2024} we will compute the conditional distribution of $\beta$ given $\bm V$ using the product rule of probability,
\begin{equation}
\label{eq:bayes}
P(\beta|\bm V, u, \beta_S) = \frac{P(\bm V, \beta|u, \beta_S)}{P(\bm V|u, \beta_S)}.
\end{equation}

The superstatistical joint distribution of velocities and inverse temperature $P(\bm V, \beta|u, \beta_S)$ is, according to \eqref{eq:super_joint} and \eqref{eq:prob_beta_kappa},
\begin{equation}
\label{eq:joint_velocities_beta}
P(\bm{V}, \beta|u, \beta_S) = \left(\frac{\tilde{m}\beta}{2\pi}\right)^{\frac{3N}{2}}\frac{1}{u\beta_S\,\Gamma\left(\frac{1}{u}\right)}
\exp\left(-\frac{\beta}{u\beta_S}\Big[1 + u\beta_S K(\bm{V})\Big]\right)\left(\frac{\beta}{u\beta_S}\right)^{\frac{1}{u}-1},
\end{equation}
thus replacing \eqref{eq:joint_velocities_beta} and \eqref{eq:multi_kappa} into \eqref{eq:bayes} yields
\begin{equation}
P(\beta|\bm V, u, \beta_S) = \frac{\Big[1+u\beta_S K(\bm V)\Big]^{\frac{3N}{2}+\frac{1}{u}}}{u\beta_S\,\Gamma\left(\frac{3N}{2}+\frac{1}{u}\right)}
\exp\left(-\frac{\beta}{u\beta_S}\Big[1 + u\beta_S K(\bm V)\Big]\right)\left(\frac{\beta}{u\beta_S}\right)^{\frac{1}{u}+\frac{3N}{2}-1}.
\end{equation}

This is an updated inverse temperature distribution, in the sense of Bayes' theorem~\cite{Sivia2006, Jaynes2003}, with respect to the prior distribution $P(\beta|u, \beta_S)$ in \eqref{eq:prob_beta_kappa} where 
we have included information about the particle velocities $\bm V$. It has also the form of a gamma distribution, but with mean
\begin{equation}
\big<\beta\big>_{\bm V, u, \beta_S} = \left(1 + \frac{3Nu}{2}\right)\frac{\beta_S}{1 + u\beta_S K(\bm V)},
\end{equation}
and relative variance
\begin{equation}
\frac{\big<(\delta \beta)^2\big>_{\bm V, u, \beta_S}}{\big<\beta\big>^2_{\bm V, u, \beta_S}} = \frac{2u}{2 + 3Nu} = \frac{1}{\kappa_N + 1},
\end{equation}
which is in general greater than zero and less than $u$, vanishing in the limit $u \rightarrow 0$ ($\kappa \rightarrow \infty$), that is, when the kappa distribution reduces to the Maxwellian, and also in the limit $N \rightarrow \infty$ for $u > 0$. In the latter case, we can approximate
\begin{equation}
\big<\beta\big>_{\bm V, u, \beta_S} \approx \frac{3N}{2K(\bm V)},
\end{equation}
and, together with the fact that $\big<(\delta \beta)^2\big>_{\bm V, u, \beta_S} \rightarrow 0$, we are led to conclude that
\begin{equation}
\label{eq:limit}
\lim_{N \rightarrow \infty} P(\beta|\bm{v}_1, \bm{v}_2, \ldots, \bm{v}_N, u, \beta_S) = \delta\big(\beta - \hat{\beta}_{\infty}\big),
\end{equation}
then agreeing with \eqref{eq:beta_no_uncertainty}, where
\begin{equation}
\label{eq:hat_beta}
\hat{\beta}_{\infty}(\bm{v}_1, \bm{v}_2, \ldots) \defeq \lim_{N \rightarrow \infty} \frac{3N}{2 K(\bm{v}_1, \bm{v}_2, \ldots, \bm{v}_N)}.
\end{equation}

In fact this means that only in the thermodynamic limit, there is a unique value of inverse temperature given the set of velocities $\bm{v}_1, \bm{v}_2, \ldots$, namely $\hat{\beta}_{\infty}$ in 
\eqref{eq:hat_beta}. It follows by taking expectation under $P(\bm V|u, \beta_S)$ on both sides of \eqref{eq:limit} and recognizing that $P(\beta|u, \beta_S)$ is independent of $N$, that
\begin{equation}
\label{eq:gravanis_pre}
P(\beta|u, \beta_S) = \left<\delta\left(\beta-\hat{\beta}_{\infty}\right)\right>_{u, \beta_S}.
\end{equation}

This is an interesting result: combined with \eqref{eq:hat_beta}, it precisely recovers the result by Gravanis \emph{et al}~\cite{Gravanis2020} on the interpretation of the superstatistical inverse 
temperature distribution as the distribution of $3N/2K$ in the thermodynamic limit, hence for clarity we can rewrite the right-hand side of \eqref{eq:gravanis_pre} as
\begin{equation}
P(\beta|u, \beta_S) = \lim_{N \rightarrow \infty} \left<\delta\Big(\beta-\frac{3N}{2K}\Big)\right>_{u, \beta_S}.
\end{equation}

Here the practical meaning of \eqref{eq:hat_beta} becomes much clear: the histogram of $3N/2K$ approaches the (invariant) inverse temperature distribution of superstatistics as the number of particles increases.
Upon rewriting the Dirac delta according to
\begin{equation}
\delta\left(\beta - \frac{3N}{2K}\right) = \frac{3N}{2\beta^2}\delta\left(K-\frac{3N}{2\beta}\right),
\end{equation}
can be written as
\begin{equation}
\label{eq:gravanis}
P(\beta|u, \beta_S) = \lim_{N \rightarrow \infty}\;\frac{3N}{2\beta^2}\, P\Big(K = \frac{3N}{2\beta}\Big|u, \beta_S\Big).
\end{equation}

It is straightforward to verify this general result in the particular case of kappa distributions. In order to do this, we replace the kappa distribution of kinetic energies for a system of $N$ particles,
\begin{equation}
P(K|u, \beta_S) = \frac{C_N(u)}{Z_N(\beta_S)}\Big[1+u\beta_S K\Big]^{-\left(\frac{1}{u}+\frac{3N}{2}\right)} \Omega_K(K; N),
\end{equation}
into the right-hand side of \eqref{eq:gravanis}, and after performing the limit operations we obtain
\begin{equation}
\begin{split}
\lim_{N \rightarrow \infty}\;\frac{3N}{2\beta^2}\, P\Big(K = \frac{3N}{2\beta}\Big|u, \beta_S\Big) & = \frac{1}{u\beta_S\,\Gamma\left(\frac{1}{u}\right)}\lim_{N \rightarrow \infty}
\frac{2\,\Gamma\left(\frac{3N}{2}+\frac{1}{u}\right)}{3N\,\Gamma\left(\frac{3N}{2}\right)} \left(\frac{3Nu\beta_S}{2\beta}\right)^{\frac{3N}{2}+1}
\Big[1+\frac{3Nu\beta_S}{2\beta}\Big]^{-\left(\frac{1}{u}+\frac{3N}{2}\right)} \\
& = \frac{1}{u\beta_S\Gamma\left(\frac{1}{u}\right)}\exp\left(-\frac{\beta}{u\beta_S}\right)\left(\frac{\beta}{u\beta_S}\right)^{\frac{1}{u}-1},
\end{split}
\end{equation}
which is precisely $P(\beta|u, \beta_S)$ as given by \eqref{eq:prob_beta_kappa}.

\section{Entropy of particles with kappa-distributed velocities}

For computing the entropy associated to a system of particles with kappa-distributed velocities, we take an approach similar to Ourabah~\cite{Ourabah2024} by using the Boltzmann-Gibbs entropy instead of any other 
generalized entropy. However, as there is uncertainty both in the microstates and the temperature of the system, we will consider the full system state as $(\bm \Gamma, \beta)$ rather than $\bm \Gamma$. Therefore, 
the entropy should be the one associated to the superstatistical joint distribution of $\bm V$ and $\beta$ in \eqref{eq:joint_velocities_beta}, namely
\begin{equation}
\label{eq:joint_entropy}
\mathcal{S}_{\beta, \bm V}(u, \beta_S) \defeq \Big<-\ln P(\bm V, \beta|u, \beta_S)\Big>_{u, \beta_S}.
\end{equation}

\begin{figure}[t!]
\begin{center}
\includegraphics[width=0.49\textwidth]{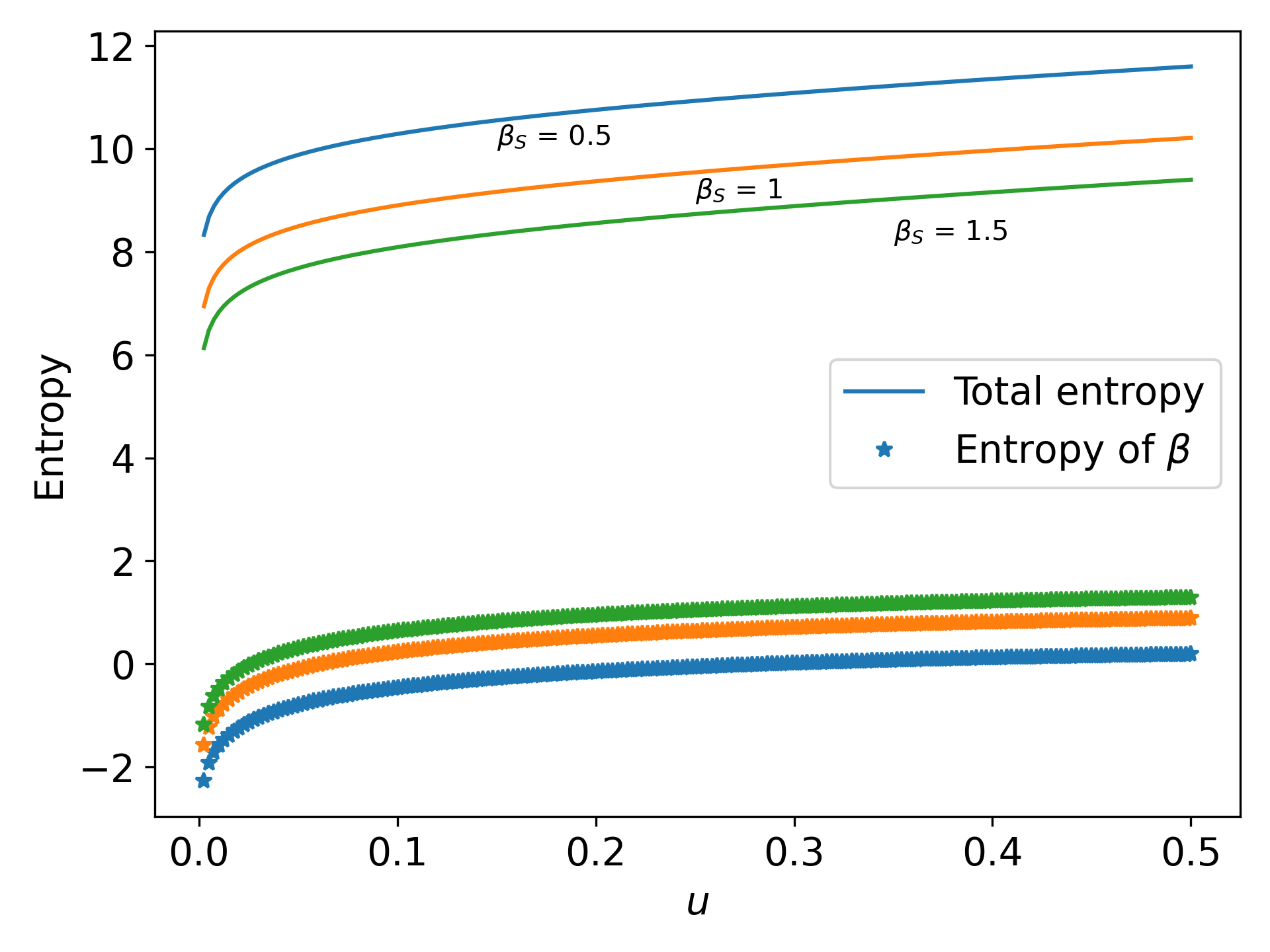}
\includegraphics[width=0.49\textwidth]{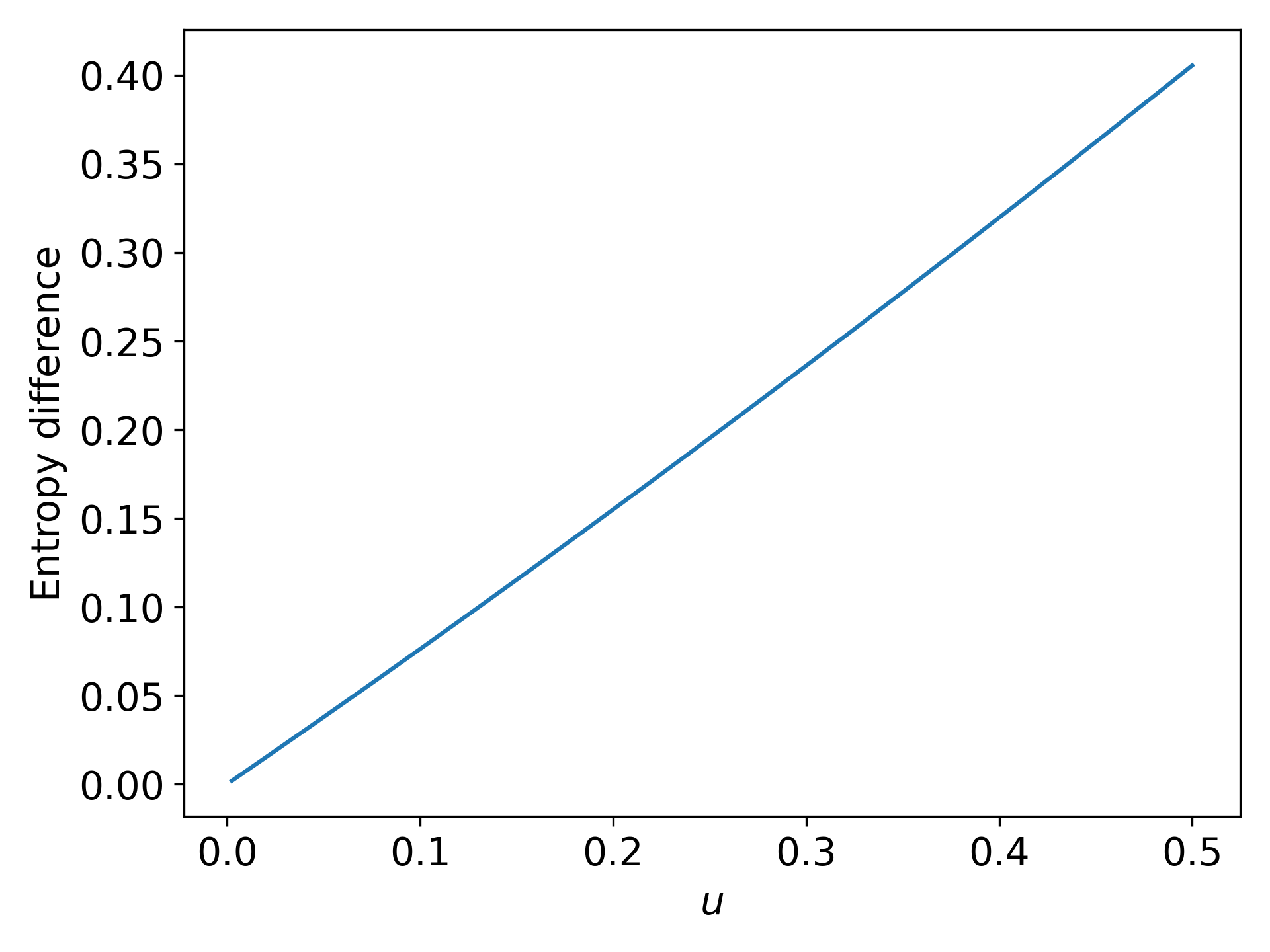}
\end{center}
\caption{Left, joint entropy $\mathcal{S}_{\beta, \bm V}$ and superstatistical entropy $\mathcal{S}_{\beta}$ as a function of $u$ for several values of $\beta_S$. Right, entropy difference $\mathcal{S}_{\text{corr}}$ 
associated to the correlation between $\beta$ and $\bm V$, as a function of $u$.}
\label{fig:entropies}
\end{figure}

\noindent
Replacing \eqref{eq:joint_velocities_beta} into \eqref{eq:joint_entropy} and simplifying the expectation values we can write
\begin{equation}
\mathcal{S}_{\beta, \bm V}(u, \beta_S) = \frac{\big<\beta\big>_{u, \beta_S}}{u\beta_S} + \big<\beta K\big>_{u, \beta_S} -\left(\frac{3N}{2}+\frac{1}{u}-2\right)\big<\ln \beta\big>_{u, \beta_S} 
+ \left(\frac{1}{u}-1\right)\ln\;(u\beta_S) + \ln \Gamma\left(\frac{1}{u}\right) -\frac{3N}{2}\ln\;\left(\frac{\tilde{m}}{2\pi}\right),
\end{equation}
where we see that $\mathcal{S}_{\beta, \bm V}$ is expressed only in terms of $\big<\beta\big>_{u, \beta_S}$, $\big<\beta K\big>_{u, \beta_S}$ and $\big<\ln \beta\big>_{u, \beta_S}$. Now, using 
\eqref{eq:beta_mean} and
\begin{equation}
\big<\beta K\big>_\beta = \frac{3N}{2}
\end{equation}
to obtain
\begin{equation}
\big<\beta K\big>_{u, \beta_S} = \left< \big<\beta K\big>_\beta \right>_{u, \beta_S} = \frac{3N}{2},
\end{equation}
together with
\begin{equation}
\big<\ln \beta\big>_{u, \beta_S} = \ln\,(u\beta_S) + \psi\left(\frac{1}{u}\right),
\end{equation}
we finally can write
\begin{equation}
\label{eq:kin_entropy}
\mathcal{S}_{\beta, \bm V}(u, \beta_S) = \mathcal{S}_{\beta}(u, \beta_S) + \mathcal{S}_{\bm V}(\beta_S) + \mathcal{S}_{\text{corr}}(u)
\end{equation}
where
\begin{equation}
\mathcal{S}_\beta (u, \beta_S) \defeq \Big<-\ln P(\beta|u, \beta_S)\Big>_{u, \beta_S} = \frac{1}{u} -F\left(\frac{1}{u}\right) + \psi\left(\frac{1}{u}\right) + \ln\,(u\beta_S)
\end{equation}
is the entropy of the superstatistical inverse temperature distribution in \eqref{eq:prob_beta_kappa},
\begin{equation}
\mathcal{S}_{\bm V}(\beta) = \Big<-\ln P(\bm V|\beta)\Big>_\beta = \frac{3N}{2}\Big(1 + \ln\;(2\pi) - \ln\;\tilde{m} - \ln \beta\Big)
\end{equation}
is the canonical entropy at inverse temperature $\beta$ and
\begin{equation}
\mathcal{S}_{\text{corr}}(u) \defeq -\frac{3N}{2}\Bigg[\psi\Big(\frac{1}{u}\Big) + \ln\,u\Bigg]
\end{equation}
is the entropy difference associated to the existence of correlations between $\beta$ and $\bm V$. These components are shown in Fig.~\ref{fig:entropies} as functions of $u$. 
The result in \eqref{eq:kin_entropy} can be directly obtained from the decomposition of the joint entropy as
\begin{equation}
\mathcal{S}_{\beta, \bm V}(u, \beta_S) = S_\beta(u, \beta_S) + \Big<\mathcal{S}_{\bm V}(\beta)\Big>_{u, \beta_S}.
\end{equation}

We see that the entropy $\mathcal{S}_{\beta, \bm V}$ is, in general, non-additive, due precisely to the first term that reflects the uncertainty in $\beta$. For instance, for two particles 
the entropy can be written as
\begin{equation}
\mathcal{S}_{\beta, \bm{v}_1, \bm{v}_2} = \mathcal{S}_{\beta, \bm{v}_1} + \mathcal{S}_{\beta, \bm{v}_2} - \mathcal{S}_\beta (u, \beta_S).
\end{equation}

\noindent
Nevertheless, in the thermodynamic limit it becomes extensive for $u > 0$, and the entropy per particle becomes
\begin{equation}
\lim_{N \rightarrow \infty} \frac{\mathcal{S}_{\beta, \bm V}(u, \beta_S)}{N} = \frac{3}{2}\Big(1 + \ln\;(2\pi) - \ln\,\tilde{m} - \ln\,\beta_S\Big) - \frac{3}{2}\left[\psi\left(\frac{1}{u}\right) + \ln u\right],
\end{equation}
which is larger than the canonical entropy per particle at $\beta = \beta_S$ for $u > 0$. The derivative of $\mathcal{S}_{\beta, \bm V}(u, \beta_S)$ with respect to $u$ is
\begin{equation}
\frac{\partial \mathcal{S}_{\beta, \bm V}(u, \beta_S)}{\partial u} = \frac{(3N-2)u + 2}{2u^3}\;\Big(\psi'\left(\frac{1}{u}\right)-u\Big)
\end{equation}
which is non-negative, thus entropy increases monotonically with $u$, as expected. On the other hand, we can evaluate the derivative
\begin{equation}
\label{eq:entropy_deriv}
\left(\frac{\partial \mathcal{S}_{\beta, \bm V}}{\partial K_S}\right)_{u, N} = \frac{\partial \mathcal{S}_{\beta, \bm V}(u, \beta_S)}{\partial \beta_S}\cdot \frac{\partial \beta_S}{\partial K_S}
= \left(\frac{3N-2}{3N}\right)\beta^*,
\end{equation}
with $\beta^*$ the most probable inverse temperature according to \eqref{eq:beta_mode}. Therefore we have
\begin{equation}
\lim_{N \rightarrow \infty} \left(\frac{\partial \mathcal{S}_{\beta, \bm V}}{\partial K_S}\right)_{u, N} = \beta^*,
\end{equation}
consistent with the fact that $\beta^*$ seems to play the role of the inverse equipartition temperature.

\section{Concluding remarks}

We have derived the multi-particle and single-particle kappa distributions from the assumption of a scale-invariant gamma distribution of inverse temperatures in superstatistics.
In light of these recent developments, the theory of superstatistics appears to be a promising candidate for an explanation of the origin of the kappa distribution in collisionless plasmas. 
Moreover, a superstatistical perspective also brings analytical tools from probability theory that allow us to simplify the computation of observables in superstatistical states. We have 
illustrated through the computation of the Pearson correlation and the mutual information between kinetic energies of two distinct particles in a kappa distribution that those energies are always 
positively correlated, and that their correlation increases with the distance to equilibrium, as measured by $u$. Furthermore, we are able to show the result by Gravanis \emph{et al} on the interpretation 
of the superstatistical temperature in kappa plasmas in a new light, as the vanishing in the thermodynamic limit of the uncertainty of the conditional distribution $P(\beta|\bm V, u, \beta_S)$ which takes 
into account all particle velocities in the system. Most of the analytical techniques developed in this work, including taking expectation over Maxwellian moments, are also applicable to other families 
of superstatistics, such as the inverse $\chi^2$ and the lognormal superstatistics. In particular, as $u$ and $\beta_S$ can also be computed for these distributions, we can employ the same invariant 
parameterization in these families.

\section*{Acknowledgments}

\noindent
Financial support from ANID FONDECYT 1220651 grant is gratefully acknowledged.

%
%
%

\begin{thebibliography}{10}

\bibitem{Antonova2008}
E.~E. Antonova and N.~O. Ermakova.
\newblock Kappa distribution functions and the main properties of auroral
  particle acceleration.
\newblock {\em Adv. Space Res.}, 42:987--991, 2008.

\bibitem{Espinoza2018}
C.~M. Espinoza, M.~V. Stepanova, P.~S. Moya, E.~E. Antonova, and J.~A.
  Valdivia.
\newblock Ion and electron kappa distribution functions along the plasma sheet.
\newblock {\em Geophysical Research Letters}, 45:6362--6370, 2018.

\bibitem{Kirpichev2020}
I.~P. Kirpichev and E.~E. Antonova.
\newblock Dependencies of kappa parameter on the core energy of kappa
  distributions and plasma parameter in the case of the magnetosphere of the
  earth.
\newblock {\em Astrophys. J.}, 891:35, 2020.

\bibitem{Eyelade2021}
A.~V. Eyelade, M.~V. Stepanova, C.~M. Espinoza, and P.~S. Moya.
\newblock On the relation between kappa distribution functions and the plasma
  beta parameter in the earth's magnetosphere: Themis observations.
\newblock {\em Astrophys. J.}, 253:34, 2021.

\bibitem{Maksimovic1997b}
M.~Maksimovic, V.~Pierrard, and J.~F. Lemaire.
\newblock A kinetic model of the solar wind with kappa distribution functions
  in the corona.
\newblock {\em Astron. Astrophys.}, 324:725--734, 1997.

\bibitem{Nicolaou2019}
G.~Nicolaou and G.~Livadiotis.
\newblock Long-term correlations of polytropic indices with kappa distributions
  in solar wind plasma near 1 au.
\newblock {\em Astrophys. J.}, 884:52, 2019.

\bibitem{ZentenoQuinteros2021}
B.~Zenteno-Quinteros, A.~F. Viñas, and P.~S. Moya.
\newblock Skew-kappa distribution functions and whistler heat flux instability
  in the solar wind: the core-strahlo model.
\newblock {\em Astrophys. J.}, 923:180, 2021.

\bibitem{Raymond2010}
J.~C. Raymond, P.~F. Winkler, W.~P. Blair, J.~Lee, and S.~Park.
\newblock Non-maxwellian h$\alpha$ profiles in tycho’s supernova remnant.
\newblock {\em Astrophys. J.}, 712:901, 2010.

\bibitem{Nicholls2017}
D.~C. Nicholls, M.~A. Dopita, R.~S. Sutherland, and L.~J. Kewley.
\newblock Electron kappa distributions in astrophysical nebulae.
\newblock In {\em Kappa Distributions}, pages 633--655. Elsevier, 2017.

\bibitem{Pierrard2010}
V.~Pierrard and M.~Lazar.
\newblock Kappa distributions: Theory and applications in space plasmas.
\newblock {\em Solar Physics}, 267:153--174, 2010.

\bibitem{Livadiotis2017}
G.~Livadiotis.
\newblock {\em Kappa distributions: Theory and applications in plasmas}.
\newblock Elsevier, 2017.

\bibitem{Lazar2021}
M.~Lazar and H.~Fichtner.
\newblock {\em Kappa distributions}.
\newblock Springer, 2021.

\bibitem{Cairns1995}
R.~A. Cairns, A.~A. Mamum, R.~Bingham, R.~Bostr{\"o}m, R.~O. Dendy, C.~M.~C.
  Nairn, and P.~K. Shukla.
\newblock Electrostatic solitary structures in non-thermal plasmas.
\newblock {\em Geophys. Res. Lett.}, 22(20):2709--2712, 1995.

\bibitem{Naim2019}
H.~Naim, I.~A. Khan, Z.~Iqbal, and G.~Murtaza.
\newblock Effect of anisotropic cairns distribution on drift magnetosonic wave.
\newblock {\em Eur. Phys. J. Plus}, 134:442, 2019.

\bibitem{Khalid2025}
M.~Khalid, S.~I.~H. Bacha, and E.~A. Elghmaz.
\newblock Ion-acoustic localized structures in a magnetized two-temperature
  electrons plasma with cairns distribution.
\newblock {\em EPL}, 152:14003, 2025.

\bibitem{Kaniadakis2001}
G.~Kaniadakis.
\newblock Non-linear kinetics underlying generalized statistics.
\newblock {\em Phys. A}, 296:405--425, 2001.

\bibitem{Lourek2016}
I.~Lourek and M.~Tribeche.
\newblock On the role of the $\kappa$-deformed kaniadakis distribution in
  nonlinear plasma waves.
\newblock {\em Phys. A}, 441:215--220, 2016.

\bibitem{Gougam2016}
L.~A. Gougam and M.~Tribeche.
\newblock Electron-acoustic waves in a plasma with a $\kappa$-deformed
  kaniadakis electron distribution.
\newblock {\em Physics of Plasmas}, 23, 2016.

\bibitem{Lopez2017b}
R.~A. López, R.~E. Navarro, S.~I. Pons, and J.~A. Araneda.
\newblock Landau damping in kaniadakis and tsallis distributed electron
  plasmas.
\newblock {\em Physics of Plasmas}, 24:102119, 2017.

\bibitem{Dum1974}
C.~T. Dum, R.~Chodura, and D.~Biskamp.
\newblock Turbulent heating and quenching of the ion sound instability.
\newblock {\em Phys. Rev. Lett.}, 32(22):1231, 1974.

\bibitem{Bissell2013}
J.~J. Bissell, C.~P. Ridgers, and R.~J. Kingham.
\newblock Super-gaussian transport theory and the field-generating thermal
  instability in laser--plasmas.
\newblock {\em New J. Phys.}, 15:025017, 2013.

\bibitem{Knapp2013}
P.~F. Knapp, D.~B. Sinars, and K.~D. Hahn.
\newblock Diagnosing suprathermal ion populations in {Z}-pinch plasmas using
  fusion neutron spectra.
\newblock {\em Phys. Plasmas}, 20:62701, 2013.

\bibitem{Klir2015}
D.~Klir, A.~V. Shishlov, V.~A. Kokshenev, P.~Kubes, A.~Yu. Labetsky, K.~Rezac,
  R.~K. Cherdizov, J.~Cikhardt, B.~Cikhardtova, G.~N. Dudkin, et~al.
\newblock Efficient generation of fast neutrons by magnetized deuterons in an
  optimized deuterium gas-puff {Z}-pinch.
\newblock {\em Plasma Phys. Control. Fusion}, 57:044005, 2015.

\bibitem{Potter1971}
D.~E. Potter and M.~G. Haines.
\newblock Non-adiabatic ions in the distribution function from self-consistent
  calculations of the plasma focus (iaea-cn-28/d-8).
\newblock {\em Plasma Physics and Controlled Nuclear Fusion Research, Volume
  I}, page 611, 1971.

\bibitem{Bernstein1972}
M.~J. Bernstein.
\newblock Neutron energy and flux distributions from a crossed-field
  acceleration model of plasma focus and {Z}-pinch discharges.
\newblock {\em Phys. Fluids}, 15:700, 1972.

\bibitem{Stygar1982}
W.~Stygar, G.~Gerdin, F.~Venneri, and J.~Mandrekas.
\newblock Particle beams generated by a 6--12.5 k{J} dense plasma focus.
\newblock {\em Nuclear Fusion}, 22:1161, 1982.

\bibitem{Vikhrev2007}
V.~V. Vikhrev and V.~D. Korolev.
\newblock Neutron generation from {Z}-pinches.
\newblock {\em Plasma Phys. Rep.}, 33:356--380, 2007.

\bibitem{Vikhrev2012}
V.~V. Vikhrev and A.~D. Mironenko-Marenkov.
\newblock On the spectrum of {Z}-pinch plasma neutrons.
\newblock {\em Plasma Phys. Rep.}, 38:225--234, 2012.

\bibitem{Soto2014}
L.~Soto, C.~Pavez, J.~Moreno, M.~J. Inestrosa-Izurieta, F.~Veloso,
  G.~Gutiérrez, J.~Vergara, A.~Clausse, H.~Bruzzone, F.~Castillo, and L.~F.
  Delgado-Aparicio.
\newblock Characterization of the axial plasma shock in a table top plasma
  focus after the pinch and its possible application to testing materials for
  fusion reactors.
\newblock {\em Physics of Plasmas}, 21:122703, 2014.

\bibitem{Pavez2015}
C.~Pavez, J.~Pedreros, A.~Tarifeño-Saldivia, and L.~Soto.
\newblock Observation of plasma jets in a table top plasma focus discharge.
\newblock {\em Physics of Plasmas}, 22:40705, 2015.

\bibitem{Pavez2022}
C.~Pavez, M.~Zorondo, J.~Pedreros, A.~Sep{\'u}lveda, L.~Soto, G.~Avaria,
  J.~Moreno, S.~Davis, B.~Bora, and J.~Jain.
\newblock New evidence about the nature of plasma filaments in plasma
  accelerators of type plasma-focus.
\newblock {\em Plasma Physics and Controlled Fusion}, 65:015003, 2022.

\bibitem{Jaynes1957}
E.~T. Jaynes.
\newblock Information theory and statistical mechanics.
\newblock {\em Phys. Rev.}, 106:620--630, 1957.

\bibitem{Tsallis1988}
C.~Tsallis.
\newblock Possible generalization of {B}oltzmann-{G}ibbs statistics.
\newblock {\em J. Stat. Phys.}, 52:479--487, 1988.

\bibitem{Tsallis2009}
C.~Tsallis.
\newblock {\em Introduction to nonextensive statistical mechanics: approaching
  a complex world}.
\newblock Springer Science \& Business Media, 2009.

\bibitem{Nauenberg2003}
M.~Nauenberg.
\newblock Critique of q-entropy for thermal statistics.
\newblock {\em Phys. Rev. E}, 67(3):036114, 2003.

\bibitem{Presse2013}
S.~Press{\'e}, K.~Ghosh, J.~Lee, and K.~A. Dill.
\newblock Nonadditive entropies yield probability distributions with biases not
  warranted by the data.
\newblock {\em Phys. Rev. Lett.}, 111:180604, 2013.

\bibitem{Presse2014}
S.~Press{\'e}.
\newblock Nonadditive entropy maximization is inconsistent with {B}ayesian
  updating.
\newblock {\em Phys. Rev. E}, 90:052149, 2014.

\bibitem{Tsallis2015}
C.~Tsallis.
\newblock Conceptual inadequacy of the {S}hore and {J}ohnson axioms for wide
  classes of complex systems.
\newblock {\em Entropy}, 17:2853--2861, 2015.

\bibitem{Jizba2019}
P.~Jizba and J.~Korbel.
\newblock Maximum entropy principle in statistical inference: case for
  non-{S}hannonian entropies.
\newblock {\em Phys. Rev. Lett.}, 122:120601, 2019.

\bibitem{Caticha2021}
A.~Caticha.
\newblock Entropy, information, and the updating of probabilities.
\newblock {\em Entropy}, 23:895, 2021.

\bibitem{Marechal2024}
P.~Maréchal, Y.~Navarrete, and S.~Davis.
\newblock On the foundations of the maximum entropy principle using {F}enchel
  duality for {S}hannon and {T}sallis entropies.
\newblock {\em Phys. Scripta}, 99:075265, 2024.

\bibitem{Beck2003}
C.~Beck and E.G.D. Cohen.
\newblock Superstatistics.
\newblock {\em Phys. A}, 322:267--275, 2003.

\bibitem{Beck2004}
C.~Beck.
\newblock Superstatistics: theory and applications.
\newblock {\em Continuum Mech. Thermodyn.}, 16:293--304, 2004.

\bibitem{Sanchez2021}
E.~S{\'a}nchez, M.~Gonz{\'a}lez-Navarrete, and C.~Caama{\~n}o-Carrillo.
\newblock Bivariate superstatistics: an application to statistical plasma
  physics.
\newblock {\em Eur. Phys. J. B}, 94:1--7, 2021.

\bibitem{Chung2020}
W.~S. Chung and H.~Hassanabadi.
\newblock Doubly superstatistics with bivariate modified dirac delta
  distribution.
\newblock {\em Physica A}, 554:124712, 2020.

\bibitem{Pachter2024}
J.~A. Pachter, Y.-J. Yang, and K.~A. Dill.
\newblock Entropy, irreversibility and inference at the foundations of
  statistical physics.
\newblock {\em Nat. Rev. Phys.}, 6:382--393, 2024.

\bibitem{Ourabah2015}
K.~Ourabah, L.~A. Gougam, and M.~Tribeche.
\newblock Nonthermal and suprathermal distributions as a consequence of
  superstatistics.
\newblock {\em Phys. Rev. E}, 91:12133, 2015.

\bibitem{Davis2019b}
S.~Davis, G.~Avaria, B.~Bora, J.~Jain, J.~Moreno, C.~Pavez, and L.~Soto.
\newblock Single-particle velocity distributions of collisionless, steady-state
  plasmas must follow superstatistics.
\newblock {\em Phys. Rev. E}, 100:023205, 2019.

\bibitem{Ourabah2020b}
K.~Ourabah.
\newblock Demystifying the success of empirical distributions in space plasmas.
\newblock {\em Phys. Rev. Research}, 2:23121, 2020.

\bibitem{Gravanis2021}
E.~Gravanis, E.~Akylas, C.~Michailides, and G.~Livadiotis.
\newblock Superstatistics and isotropic turbulence.
\newblock {\em Phys. A}, 567:125694, 2021.

\bibitem{Dixit2013}
P.~D. Dixit.
\newblock A maximum entropy thermodynamics of small systems.
\newblock {\em J. Chem. Phys.}, 138:184111, 2013.

\bibitem{Dixit2015}
P.~D. Dixit.
\newblock Detecting temperature fluctuations at equilibrium.
\newblock {\em Phys. Chem. Chem. Phys.}, 17:13000--13005, 2015.

\bibitem{Herron2021}
L.~Herron and P.~Dixit.
\newblock Thermal behavior of small magnets.
\newblock {\em J. Stat. Mech.: Theor. Exp.}, 2021:033207, 2021.

\bibitem{Jizba2010}
P.~Jizba and H.~Kleinert.
\newblock Superstatistics approach to path integral for a relativistic
  particle.
\newblock {\em Phys. Rev. D}, 82:085016, 2010.

\bibitem{Ayala2018}
A.~Ayala, M.~Hentschinski, L.~A. Hernández, M.~Loewe, and R.~Zamora.
\newblock Superstatistics and the effective {QCD} phase diagram.
\newblock {\em Phys. Rev. D}, 98:114002, 2018.

\bibitem{Ourabah2019}
K.~Ourabah, E.~M.~Barboza Jr., E.~M.~C. Abreu, and J.~A. Neto.
\newblock Superstatistics: Consequences on gravitation and cosmology.
\newblock {\em Phys. Rev. D}, 100:103516, 2019.

\bibitem{Chen2008}
L.~L. Chen and C.~Beck.
\newblock A superstatistical model of metastasis and cancer survival.
\newblock {\em Phys. A}, 387:3162--3172, 2008.

\bibitem{Denys2016}
M.~Denys, T.~Gubiec, R.~Kutner, M.~Jagielski, and H.~E. Stanley.
\newblock Universality of market superstatistics.
\newblock {\em Phys. Rev. E}, 94:042305, 2016.

\bibitem{Bogachev2017}
M.~I. Bogachev, O.~A. Markelov, A.~R. Kayumov, and A.~Bunde.
\newblock Superstatistical model of bacterial {DNA} architecture.
\newblock {\em Sci. Rep.}, 7:43034, 2017.

\bibitem{Schafer2018}
B.~Schäfer, C.~Beck, K.~Aihara, D.~Witthaut, and M.~Timme.
\newblock Non-{G}aussian power grid frequency fluctuations characterized by
  {L}évy-stable laws and superstatistics.
\newblock {\em Nat. Energy}, 3:119--126, 2018.

\bibitem{Costa2022}
M.~O. Costa, R.~Silva, and D.~H. A.~L. Anselmo.
\newblock Superstatistical and {DNA} sequence coding of the human genome.
\newblock {\em Phys. Rev. E}, 106:064407, 2022.

\bibitem{Sanchez2025}
E.~Sánchez.
\newblock Gamma-superstatistics and complex time series analysis.
\newblock {\em Phys. Rev. E}, 112:014118, 2025.

\bibitem{Davis2023e}
S.~Davis, G.~Avaria, B.~Bora, J.~Jain, J.~Moreno, C.~Pavez, and L.~Soto.
\newblock Kappa distribution from particle correlations in nonequilibrium,
  steady-state plasmas.
\newblock {\em Phys. Rev. E}, 108:065207, 2023.

\bibitem{Sattin2006}
F.~Sattin.
\newblock Bayesian approach to superstatistics.
\newblock {\em Eur. Phys. J. B}, 49:219--224, 2006.

\bibitem{Abe2014b}
S.~Abe.
\newblock Fokker-planck theory of nonequilibrium systems governed by
  hierarchical dynamics.
\newblock {\em Found. Phys.}, 44:175--182, 2014.

\bibitem{Davis2023b}
S.~Davis.
\newblock Superstatistics and the fundamental temperature of steady states.
\newblock {\em AIP Conf. Proc.}, 2731:30006, 2023.

\bibitem{Davis2022}
S.~Davis.
\newblock Fluctuating temperature outside superstatistics: Thermodynamics of
  small systems.
\newblock {\em Phys. A}, 589:126665, 2022.

\bibitem{Davis2022b}
S.~Davis.
\newblock A classification of nonequilibrium steady states based on temperature
  correlations.
\newblock {\em Phys. A}, 608:128249, 2022.

\bibitem{Livadiotis2019}
G.~Livadiotis.
\newblock Theoretical aspects of {H}amiltonian kappa distributions.
\newblock {\em Physica Scripta}, 94:105009, 2019.

\bibitem{Nicolaou2020c}
G.~Nicolaou and G.~Livadiotis.
\newblock Statistical uncertainties of space plasma properties described by
  kappa distributions.
\newblock {\em Entropy}, 22:541, 2020.

\bibitem{Livadiotis2022}
G.~Livadiotis and D.~J. McComas.
\newblock Physical correlations lead to kappa distributions.
\newblock {\em The Astrophysical Journal}, 940:83, 2022.

\bibitem{CoverThomas2006}
T.~M. Cover and J.~A. Thomas.
\newblock {\em Elements of Information Theory}.
\newblock John Wiley and Sons, 2006.

\bibitem{Davis2024}
S.~Davis.
\newblock A superstatistical measure of distance from canonical equilibrium.
\newblock {\em J. Phys. A: Math. Theor.}, 57:295004, 2024.

\bibitem{Abramowitz1972}
M.~Abramowitz and I.~Stegun.
\newblock {\em Handbook of {M}athematical {F}unctions}.
\newblock Dover, 1972.

\bibitem{Davis2018}
S.~Davis and G.~Gutiérrez.
\newblock Temperature is not an observable in superstatistics.
\newblock {\em Phys. A}, 505:864--870, 2018.

\bibitem{Sivia2006}
D.~Sivia and J.~Skilling.
\newblock {\em Data analysis: a Bayesian tutorial}.
\newblock OUP Oxford, 2006.

\bibitem{Jaynes2003}
E.~T. Jaynes.
\newblock {\em Probability Theory: The Logic of Science}.
\newblock Cambridge University Press, 2003.

\bibitem{Gravanis2020}
E.~Gravanis, E.~Akylas, and G.~Livadiotis.
\newblock Physical meaning of temperature in superstatistics.
\newblock {\em EPL}, 130:30005, 2020.

\bibitem{Ourabah2024}
K.~Ourabah.
\newblock Superstatistics from a dynamical perspective: Entropy and relaxation.
\newblock {\em Phys. Rev. E}, 109:014127, 2024.

\end{thebibliography}

\end{document}